\documentclass [10pt, a4paper]{article}
\usepackage {etex}

\usepackage [english]{babel}
\usepackage [T1]{fontenc}
\usepackage [utf8] {inputenc}
\usepackage {lmodern}
\usepackage [left=2cm, right=2cm, top=2cm, bottom=2cm] {geometry}
\usepackage {amsfonts}
\usepackage {amsmath}	
\usepackage {amssymb}
\usepackage {amsxtra}
\usepackage {wasysym}
\usepackage {mathtools}
\usepackage {calrsfs}
\usepackage {dsfont}
\usepackage {fancyhdr}
\usepackage {graphicx}
\usepackage [amsthm,thmmarks,amsmath]{ntheorem} 
\usepackage {pstricks}
\usepackage {pst-plot}
\usepackage {pst-vue3d}
\usepackage {exscale}
\usepackage {cite} 
\usepackage {setspace}
\usepackage {color}
\usepackage [labelfont=bf,labelsep=endash,width=.8\textwidth]{caption}
\usepackage [bottom]{footmisc}
\usepackage [all,cmtip]{xy}
\usepackage {hyperref}
\usepackage [affil-it, blocks]{authblk}

\allowdisplaybreaks
\singlespace
\input diagxy

\makeatletter
\@addtoreset{equation}{section}
\renewcommand\theequation{\thesection.\@arabic\c@equation}
\renewcommand\thefigure{\thesection.\@arabic\c@figure}
\renewcommand\thetable{\thesection.\@arabic\c@table}
\makeatother

\DeclareFontFamily{OT1}{pzc}{}
\DeclareFontShape{OT1}{pzc}{m}{it}{<-> s * [1.30] pzcmi7t}{}
\DeclareMathAlphabet{\mathpzc}{OT1}{pzc}{m}{it}

\DeclareMathOperator{\R}{\mathds R}
\DeclareMathOperator{\C}{\mathds C}
\DeclareMathOperator{\N}{\mathds N}
\DeclareMathOperator{\M}{\mathds M}

\DeclareMathOperator{\supp}{supp}
\DeclareMathOperator{\ssupp}{sing\ supp}

\DeclareMathOperator{\id}{id}

\DeclareMathOperator{\cD}{\mathcal D}
\DeclareMathOperator{\WF}{\textup WF}

\DeclareMathOperator{\cE}{\mathcal E}
\DeclareMathOperator{\Char}{\textup Char}
\DeclareMathOperator{\rank}{\textup rank}
\DeclareMathOperator{\usc}{\textup sc}
\DeclareMathOperator{\im}{\textup im}
\DeclareMathOperator{\Sol}{\textup Sol}
\DeclareMathOperator{\cW}{\mathcal W}
\DeclareMathOperator{\sinc}{\textup sinc}

\newcommand{\BIGOP}[1]{\mathop{\mathchoice
{\raise-0.22em\hbox{\huge $#1$}}
{\raise-0.05em\hbox{\Large $#1$}}{\hbox{\large $#1$}}{#1}}}

\renewcommand{\div}{\mathrm{div}}

\newtheoremstyle{breakdef}%
  {\item[\rlap{\vbox{\normalfont\bfseries\hbox{\llap{##2}\hskip\labelsep
          ##1:}\hbox{\\[0.1cm]}}}]}%
  {\item[\rlap{\vbox{\normalfont\bfseries\hbox{\llap{##2}\hskip\labelsep
          ##1 (##3):}\hbox{\\[0.1cm]}}}]}
\newtheoremstyle{breaksatz}%
  {\item[\rlap{\vbox{\normalfont\normalsize\bfseries\hbox{\llap{##2}\hskip\labelsep
          ##1:}\hbox{\\[0.1cm]}}}]}%
  {\item[\rlap{\vbox{\normalfont\normalsize\bfseries\hbox{\llap{##2}\hskip\labelsep
          ##1 (##3):}\hbox{\\[0.1cm]}}}]}
\newtheoremstyle{breaklem}%
  {\item[\rlap{\vbox{\normalfont\normalsize\bfseries\hbox{\llap{##2}\hskip\labelsep
          ##1:}\hbox{\\[0.1cm]}}}]}%
  {\item[\rlap{\vbox{\normalfont\normalsize\bfseries\hbox{\llap{##2}\hskip\labelsep
          ##1 (##3):}\hbox{\\[0.1cm]}}}]}
\newtheoremstyle{breakprop}%
  {\item[\rlap{\vbox{\normalfont\normalsize\bfseries\hbox{\llap{##2}\hskip\labelsep
          ##1:}\hbox{\\[0.1cm]}}}]}%
  {\item[\rlap{\vbox{\normalfont\normalsize\bfseries\hbox{\llap{##2}\hskip\labelsep
          ##1 (##3):}\hbox{\\[0.1cm]}}}]}
\newtheoremstyle{breakbem}%
  {\item[\rlap{\vbox{\hbox{\hskip\labelsep\normalfont\bfseries
          ##1 ##2:}\hbox{\\[0.1cm]}}}]}%
  {\item[\rlap{\vbox{\hbox{\hskip\labelsep\normalfont\bfseries
          ##1 ##2 (##3):}\hbox{\\[0.1cm]}}}]}
\newtheoremstyle{breakbsp}%
  {\item[\rlap{\vbox{\hbox{\hskip\labelsep\normalfont\bfseries
          ##1 ##2:}\hbox{\\[0.2cm]}}}]}%
  {\item[\rlap{\vbox{\hbox{\hskip\labelsep\normalfont\bfseries
          ##1 ##2 (##3):}\hbox{\\[0.2cm]}}}]}
\newtheoremstyle{breakkor}%
  {\item[\rlap{\vbox{\hbox{\hskip\labelsep\normalfont\bfseries
          ##1 ##2:}\hbox{\\[0.1cm]}}}]}%
  {\item[\rlap{\vbox{\hbox{\hskip\labelsep\normalfont\bfseries
          ##1 ##2 (##3):}\hbox{\\[0.1cm]}}}]}
\newtheoremstyle{proof}%
  {\item[\rlap{\vbox{\hbox{\hskip\labelsep\normalfont\bfseries
          \underline{##1:}}\hbox{\\[0.1cm]}}}]}%
  {\item[\rlap{\vbox{\hbox{\hskip\labelsep\normalfont\bfseries
          \underline{##1 (##3):}}\hbox{\\[0.1cm]}}}]}
\theoremstyle{breakkor} 
\newtheorem{definition}{Definition}[section]
\theoremstyle{breakkor}
\newtheorem{theorem}[definition]{Theorem}
\theoremstyle{breakkor}
\newtheorem{lemma}[definition]{Lemma}
\theoremstyle{breakkor}
\newtheorem{proposition}[definition]{Proposition}
\theoremstyle{breakkor}
\newtheorem{corollary}[definition]{Corollary}
\theoremstyle{breakkor}
\theorembodyfont{\normalfont}
\newtheorem{remark}[definition]{Remark}
\newtheorem{example}[definition]{Example}
\theoremstyle{Proof}
\theoremsymbol{\fbox{}}

\begin{document}

\lhead[\thepage]{\rightmark} 
\chead[]{}
\rhead[\leftmark]{\thepage}
\lfoot[]{}
\cfoot[]{}
\rfoot[]{}

\thispagestyle{empty}

\title{\Large\bf The microlocal spectrum condition, initial value formulations and background independence}
\author[]{Alexander Stottmeister\thanks{{\sf 
\hspace{0.1cm}alexander.stottmeister@gravity.fau.de}}\ }
\author[]{Thomas Thiemann\thanks{{\sf 
\hspace{0.1cm}thomas.thiemann@gravity.fau.de}}}
\affil[]{Institut für Quantengravitation, Lehrstuhl für Theoretische Physik III, \\ Friedrich-Alexander-Universtität Erlangen-Nürnberg, \\ Staudtstraße 7/B2, D-91058 Erlangen, Germany}

\maketitle

\begin{abstract}
We analyze implications of the microlocal spectrum/Hadamard condition for states in a (linear) quantum field theory on a globally hyperbolic spacetime $M$ in the context of a (distributional) initial value formulation. More specifically, we work in $3+1$-split $M\cong\R\times\Sigma$ and give a bound, independent of the spacetime metric, on the wave front sets of the initial data for a quasi-free Hadamard state in a quantum field theory defined by a normally hyperbolic differential operator $P$ acting in a vector bundle $E\stackrel{\pi}{\rightarrow}M$. This aims at a possible way to apply the concept of Hadamard states within approaches to quantum field theory/gravity relying on a Hamiltonian formulation, potentially without a (classical) background metric $g$. 
\end{abstract}

\tableofcontents

\section{Introduction}
Quantum field theory on curved spacetimes is nowadays a well-developed subject, which allows for the rigorous treatment of perturbative quantization of classical field theories on curved spacetimes  (see \cite{FredenhagenPerturbativeAlgebraicQuantum} for a recent review), including scalar fields, Dirac fields, Yang-Mills fields, and even the treatment of perturbative quantum gravity in a locally covariant fashion \cite{BrunettiQuantumGravityFrom}. At the basis of this approach are the linear(ized) field theories and their quantum theories, which are probably the most studied examples of quantum field theories. In the framework of algebraic quantum field theory, the concept of Hadamard states for linear quantum fields plays an important role (see e.g. \cite{WaldQuantumFieldTheory, BrunettiMicrolocalAnalysisAnd, HollandsExistenceOfLocal, BrunettiTheGenerallyCovariant}). These states replace the concept of vacuum state in a locally covariant manner by mimicking the short-distance behavior of the latter in purely spacetime geometric fashion.  It is know that there are sufficiently many of these states on arbitrary globally hyperbolic spacetimes (cf. \cite{FullingSingularityStructureOf2, JunkerHadamardStatesAdiabatic, GerardConstructionOfHadamard}). A particularly elegant characterization of Hadamard states is the so-called microlocal spectrum condition \cite{RadzikowskiMicrolocalApproachTo}, which is a prescription for the wave front set of the associated two-point function(al), and can be interpreted as a remnant of the spectrum condition in quantum field theory on Minkowski space.\\
In this article, we analyze the relation between the microlocal spectrum condition and the initial value formulation of the quantum field theory with the aim to extract a condition that is manifestly independent of the spacetime metric. Such a condition could be used as a filter for physically interesting states in the matter sector of approaches to quantum gravity, where a (classical) spacetime metric is not available, e.g. loop quantum gravity. Furthermore, our analysis provides a first step to elucidate the structures that need to be present in a theory of quantum gravity coupled to matter, such that quantum field theory on curved spacetime can be extracted in a semi-classical limit. Potential candidates for a semi-classical analysis of loop quantum gravity in this regard are the deparametrizing models (see \cite{GieselScalarMaterialReference} for an overview). Notably, the concept of adiabatic vacua, which is related to that of Hadamard states \cite{LuedersLocalQuasiequivalenceAnd, JunkerAdiabaticVacuumStates}, has already been applied in the framework of loop quantum cosmology, a loop quantization of symmetry reduced models, to treat cosmological perturbations \cite{AgulloExtensionOfThe, AgulloThePreInflationary}.
\\[0.5cm]
The organization of the article is as follows:\\[0.1cm]
The main part of the article is section \ref{sec:linearqft}. It deals with the quantum theory of linear field theories in the context locally covariant quantum field theory \cite{BrunettiTheGenerallyCovariant}, and the important notion of Hadamard states, which are characterized by prescription for the wave front set of the two-point correlation function(al). In the first subsection \ref{sec:initialvalueproblem}, beside recalling important results about linear field theories on Lorentzian manifolds, we prove, building on work by Dimock \cite{DimockAlgebrasOfLocal}, a theorem on the initial value problem for generalized wave equations with distributional initial data. In subsection \ref{sec:microlocalspectrumcondition}, we discuss the microlocal spectrum/Hadamard condition, and prove the main theorem of this article: A bound on the wave front sets of the initial data for a quasi-free Hadamard state of a linear quantum field theory, which is independent of the spacetime metric.\\
In section \ref{sec:remarks}, we discuss the wave front sets’ bound for the initial data in view of dynamical aspects of the microlocal spectrum condition and available construction procedures for Hadamard states. Furthermore, we outline how the wave front sets’ bound could be applied as an a priori condition for semi-classical states of quantum matter fields in background independent theories like loop quantum gravity.\\
Section \ref{sec:app} provides an appendix with some essential material from the theory of distributions and their wave front sets.\\[0.2cm]
Let us fix some notation:\\[0.1cm]
Throughout the article, $(M,g)$, or $M$ for short, denotes spacetime, i.e. a globally hyperbolic, time-/space-oriented, Hausdorff, second-countable, $\sigma$-compact ($C^{\infty}$-) manifold ($\dim(M)=m<\infty$). The metric-induced volume form on $M$ is $dV_{g}$. A Cauchy surface for $M$ is called $\Sigma$, i.e. $M\cong\R\times\Sigma$. The induced volume form on $\Sigma$ is $dA_{g}$. For the causal future/past of a subset $K\subset M$, we use the usual notation $J_{\pm}(K)$ ($J(K):=J_{+}(K)\cup J_{-}(K)$). $K\sqsubset M$ indicates a compact subset. $E\stackrel{\pi}{\rightarrow}M$, or simply $E$, is a finite dimensional, (real) vector bundle over $M$ ($\rank(E)=e$), and $E^{*}$ its dual. If we have two such vector bundles $E,E'$, we denote the exterior tensor product over $M\times M'$ by $E\boxtimes E'$, and the interior tensor product, for $M=M'$, over $M$ by $E\otimes E'$. If we do not specify a connection in the tangent bundles $TM$, $T\Sigma$, these are given resp. induced by the Levi-Civita connection of $g$.\\[0.1cm]
The functional spaces, we frequently use, are:
\begin{itemize}
	\item[1.] The compactly supported, smooth functions on $M$ or $\Sigma$, and their distributional duals:
	\begin{center}
	$\cD(M),\ \cD(\Sigma)$ and $\cD'(M),\ \cD'(\Sigma)$.
	\end{center}
	\item[2.] The smooth functions on $M$ or $\Sigma$, and their distributional duals:
	\begin{center}
	$\cE(M),\ \cE(\Sigma)$ and $\cE'(M),\ \cE'(\Sigma)$.
	\end{center}
	\item[3.] The smooth functions with ``spacelike compact'' support on $M$, and their dual: 
	\begin{center}
	$\cE_{\usc}(M)$ and $\cE_{\usc}(M)$ (cf. \cite{BaerWaveEquationsOn})\footnote{$\cE_{\textup{sc}}(M)$ are the smooth functions on $M$ which are ``spacelike compact'', i.e. if $f\in\cE_{\textup{sc}}(M)$ there exists a compact set $K\sqsubset M$ s.t. $\supp(f)\subset J(K)$.}.
	\end{center}
	\item[4.] The generalizations of these spaces to sections in a vector bundle $E$ over $M$ or its restriction $E_{\Sigma}$ to $\Sigma$: 
	\begin{center}
	$\cD(M,E)$, $\cD'(M,E^{*})$ etc. (cf. \cite{BaerWaveEquationsOn}).
	\end{center}
	\item[5.] Distribution spaces with specified wave front sets:
	\begin{center}
	$\cD'_{\Gamma}$, $\cE'_{\Lambda}$,
	\end{center}
	where $\Gamma, \Lambda$ are conical subsets of $T^{*}M$ or $T^{*}\Sigma$ (see definition \ref{def:restrictedwavefrontsetdist}, cf. \cite{HoermanderFourierIntegralOperators, DabrowskiFunctionalPropertiesOf}).
\end{itemize}
The embeddings of the type $\cD(M,E)\hookrightarrow\cD'(M,E)$ are understood by means of the volume form $dV_{g}$ and the fibre metric $g_{E}$, i.e.
\begin{align}
\label{eq:functionembedding}
\forall f'\in\cD(M,E):\ (f,f') & :=\int_{M}g_{E}(f,f')dV_{g},
\end{align}
associates a unique distribution to every $f\in\cD(M,E)$. All spaces will be equipped with one of their usual topologies. Thus, we refrain from restating the various definitions and refer the interested reader to the appendix and references.
Let us also issue a word of caution regarding the notions of continuity and sequential continuity: In general, sequentially continuous maps between locally convex topological vector spaces are not necessarily continuous in the topological sense. Although, equality of the concepts holds for bornological topologies (cf. \cite{KrieglTheConvenientSetting}), it may fail for non-bornological spaces like $D'_{\Gamma},\ \Gamma$ a closed, but open at the tip, cone (cf. \cite{DabrowskiFunctionalPropertiesOf}). In this article, we restrict ourselves to the simpler case of sequential continuity.

\section{Linear quantum fields in curved spacetimes}
\label{sec:linearqft}
We start this section by a brief outline of some essential facts for the understanding of linear quantum fields in curved space time and our analysis of the microlocal spectrum/Hadamard condition (cf. \cite{DimockAlgebrasOfLocal, FriedlanderTheWaveEquation, BaerWaveEquationsOn, BrunettiTheGenerallyCovariant}). We conclude the first subsection \ref{sec:initialvalueproblem} by proving that the distributional initial value problem for generalized wave equations can be considered well-posed. After this, we proceed to the discussion of the microlocal spectrum/Hadamard condition for the quantum theory, and prove the main theorem of the article

\subsection{The initial value formulation for generalized wave equations}
\label{sec:initialvalueproblem}
Let us consider a spacetime $M$, and a vector bundle $E$ on $M$ equipped with a (non-degenerate) fibre metric $g_{E}$. The fibre metric $g_{E}$ provides an identification of $E$ and $E^{*}$, which we will use freely. Global hyperbolicity implies the existence of a $3+1$-split of spacetime, $M\cong\R\times S$ in the $C^{\infty}$-sense (cf. \cite{BernalSmoothnessOfTime}), and we have a well-posed initial value problem (cf. \cite{BaerWaveEquationsOn}), with initial data in $\cD(\Sigma,E_{\Sigma})$, for generalized wave equations
\begin{align}
\label{eq:waveequation}
Pu & =0,\ u\in\cE(M,E),
\end{align}
where $P:\cE(M,E)\rightarrow\cE(M,E)$ is a formally self-adjoint, normally hyperbolic differential operator, i.e. 
\begin{align}
\label{eq:formalselfadjointness}
\int_{M}g_{E}(Pu,v)dV_{g}=\int_{M}g_{E}(u,Pv)dV_{g},
\end{align}
and the principal symbol of $P$ is given by the spacetime metric $g$:
\begin{align}
\label{eq:normhypop}
P & = g^{ij}(x)\frac{\partial^{2}}{\partial x^{i}\partial x^{j}} + a^{k}(x)\frac{\partial}{\partial x^{k}} + b(x)
\end{align}
in local coordinates $x=(x^{1},...,x^{m})$ on $U\subset M$ subordinate to a local trivialization $E_{|U}\cong U\times\R^{e}$ with matrix valued coefficients $a,b:U\rightarrow\R^{e}$.
Moreover, there exist unique advanced and retarded fundamental solutions
\begin{align}
\label{eq:advancedretarded}
& G^{\pm}:\cD(M,E)\longrightarrow\cE_{\textup{sc}}(M,E), \\
& P\circ G^{\pm} = \id_{\cD(M,E)},\ G^{\pm}\circ P_{\cD(M,E)} = \id_{\cD(M,E)}\\
& \forall f\in\cD(M,E):\ \supp(G^{\pm}(f))\subset J^{M}_{\pm}(\supp(f)),
\end{align}
and we may write (cp. \cite{DimockAlgebrasOfLocal})
\begin{align}
\label{eq:solutionfromdata}
f = (G'\circ(\iota^{*})')(f_{1})-(G'\circ(\nu^{*})')(f_{0}),\ f_{0},f_{1}\in\cD(\Sigma,E_{\Sigma}^{*})\cong\cD(\Sigma,E_{\Sigma})\subset\cD'(\Sigma,E^{*}_{\Sigma}),
\end{align}
where $\iota:\Sigma\hookrightarrow M$ is the inclusion of the Cauchy surface, $G':\cE_{\usc}'(M,E^{*})\rightarrow\cD'(M,E^{*})$ is the (formal) adjoint of the causal propagator $G=G^{-}-G^{+}$, and $(\iota^{*})',(\nu^{*})':\cD'(\Sigma,E_{\Sigma}^{*})\rightarrow\cE'_{sc}(M,E^{*})$ denote the adjoints of the maps
\begin{align}
\label{eq:restrictionmaps}
\iota^{*}:\cE_{\usc}(M,E)\rightarrow\cD(\Sigma,E_{\Sigma}),\ &\ u\mapsto u_{|\Sigma}=\iota^{*}u \\[0.2cm] \nonumber
\nu^{*}:\cE_{\usc}(M,E)\rightarrow\cD(\Sigma,E_{\Sigma}),\ &\ u\mapsto(\nabla_{n}u)_{|\Sigma}=\iota^{*}(\nabla_{n}u),
\end{align}
where $n\in\cE(\Sigma,TM_{\Sigma})$ denotes the timelike, future oriented, unit normal to $\Sigma$, and $\nabla$ is the unique $P$-compatible connection in $E$ (cf. \cite{BaumNormallyHyperbolicOperators, BaerWaveEquationsOn}). We notice that the adjoints of the advanced and retarded fundamental solutions satisfy $G^{\pm}=(G^{\mp})'$, because of the formal self-adjointness of $P$. From \cite{DimockAlgebrasOfLocal, BaerWaveEquationsOn}, we know that the restrictions
\begin{align}
\label{eq:restrictionsolutionfromdata}
G'\circ(\iota^{*})',\ G'\circ(\nu^{*})':\cD(\Sigma,E^{*}_{\Sigma})\subset\cD'(\Sigma,E^{*}_{\Sigma})\rightarrow\cE_{\textup{sc}}(M,E^{*})\subset\cD'(M,E^{*})
\end{align}
are (sequentially) continuous maps, and one finds the identities
\begin{align}
\label{eq:initialdatasymplecticform}
\iota^{*}\circ G\circ(\iota^{*})'=0,\ &\ \iota^{*}\circ G\circ(\nu^{*})'= -\id_{\cD(\Sigma,E_{\Sigma}) }, \\[0.1cm] \nonumber
\nu^{*}\circ G\circ(\iota^{*})'=\id_{\cD(\Sigma,E_{\Sigma}) },\ &\ \nu^{*}\circ G\circ(\nu^{*})'=0, \\[0.25cm]
\label{eq:propagatoridentity}
G = G\circ(\iota^{*})'\circ\nu^{*}\circ G-&G\circ(\nu^{*})'\circ\iota^{*}\circ G,
\end{align}
and a short exact sequence of (sequentially) continuous maps
\begin{align}
\label{eq:shortexact}
\xymatrix{
0 \ar[r] & \cD(M,E) \ar[r]^{P} & \cD(M,E) \ar[r]^{G} & \cE_{\usc}(M,E) \ar[r]^{P} & \cE_{\usc}(M,E).
}
\end{align}
Furthermore, it follows from the results of \cite{DimockAlgebrasOfLocal}, and \eqref{eq:shortexact}, that \eqref{eq:solutionfromdata} can be utilized to construct solutions with distributional initial data $u_{0},u_{1}\in\cD'(\Sigma,E_{|\Sigma}^{*})$, i.e.
\begin{align}
\label{eq:distsolutionfromdata}
u = (G'\circ(\iota^{*})')(u_{1})-(G'\circ(\nu^{*})')(u_{0})\in\cD'(M,E^{*}).
\end{align}
\begin{remark}
\label{rem:solutionwavefrontset}
Equation \eqref{eq:distsolutionfromdata} admits an important refinement, because there exists a strong constraint on the wave front set of any distributional solution $u\in\cD'(M,E^{*})$ to a linear partial differential equation $Pu=0$ (see theorem \ref{thm:characteristicwavefrontset} ,cf. \cite{HoermanderTheAnalysisOf1}):
\begin{align}
\label{eq:constrainedwavefrontsetforsolutions}
\WF(u)\subset\Char P.
\end{align}
 The conical subset $\Char P\subset T^{*}M\setminus\{0\}$, called the \textit{characteristic set of} $P$, is defined in theorem \ref{thm:characteristicwavefrontset} of the appendix. The definition of the wave front set of a distribution can be found in the appendix (see definition \ref{def:wavefrontset}), as well. We conclude that \eqref{eq:distsolutionfromdata} can be replaced by
\begin{align}
\label{eq:distsolutionfromdatawavefrontset}
u = (G'\circ(\iota^{*})')(u_{1})-(G'\circ(\nu^{*})')(u_{0})\in\cD'_{\Char P}(M,E^{*}).
\end{align}
\end{remark}
What is not achieved in \cite{DimockAlgebrasOfLocal}, although one finds a contrary statement in \cite{JunkerHadamardStatesAdiabatic}, is an answer to the questions, which distributional solutions $u\in\cD'(M,E^{*})$ arise in this way, and in which sense the initial value problem can be considered well-posed for initial data in $\cD'(\Sigma,E_{\Sigma}^{*})$. A (partial) answer to these questions can be given in form of the following theorem.
\begin{theorem}[The distributional initial value problem]
\label{thm:distinitialvalueproblem}
Let $E$ be a vector bundle over a globally hyperbolic spacetime $M$, equipped with a non-degenerate fibre metric $g_{E}$. Furthermore, let $P$ be a formally self-adjoint, normally hyperbolic operator acting in $E$. Then, given $u_{0},u_{1}\in\cD'(\Sigma,E^{*}_{\Sigma})$, there exists a unique, proper, distributional solution $u\in\cD'_{\Char P}(M,E^{*})$ to the equation $Pu=0$, s.t. $u_{|\Sigma}=u_{0},\ (\nabla_{n}u)_{|\Sigma}=u_{1}$. Here, a distributional solution $u$ is called proper (cp. \cite{LindbladCounterexamplesToLocal}), if it can be approximated by a sequence of regular solutions $\{u_{j}\}_{j=1}^{\infty}\subset\cE(M,E^{*})$ in the (weak) topology of $\cD'_{\Char P}(M,E^{*})$, i.e.
\begin{align}
\label{eq:propersolution}
\forall f\in\cD(M,E):\ (u,f) & = \lim_{j\rightarrow\infty}(u_{j},f),\\ \nonumber
\forall j:\ Pu_{j} & = 0.
\end{align}
Moreover, the map
\begin{align}
\label{eq:solutionmap}
\cD'(\Sigma,E_{\Sigma}^{*})^{\oplus 2}\rightarrow\cD'_{\Char P}(M,E^{*})
\end{align}
sending $(u_{0},u_{1})$ to the solution $u$, s.t. $Pu=0,\ u_{|\Sigma}=u_{0},\ (\nabla_{n}u)_{|\Sigma}=u_{1}$, is (sequentially) continuous.
\end{theorem}
Before we start the proof of this theorem, we state useful results concerning a generalized Green's identity and the continuity of some of the maps introduced above.
\begin{lemma}[Green's identity for normally hyperbolic differential operators, cf. \cite{BaerWaveEquationsOn}]
\label{lem:normhypgreensidentity}
Let $P:\cE(M,E)\rightarrow\cE(M,E)$ be normally hyperbolic, and $\nabla$ be the unique $P$-compatible connection in $E$\footnote{$\nabla$ induces a connection in $E^{*}$, which we denote by the same symbol.}. Then, we have for every $u\in\cE(M,E^{*})$ and $f\in\cE(M,E)$ the identity
\begin{align}
\label{eq:normhypgreensidentity}
(u,Pf)-(P^{*}u,f) & = \div_{g}(W),
\end{align}
where $W\in\cE(M,TM)$ is defined by
\begin{align}
\label{eq:greensboundaryterm}
g(W,X) & = (\nabla_{X}u,f)-(u,\nabla_{X}f),\ \forall X\in\cE(M,E).
\end{align}
Here, $\div_{g}$ denotes the divergence operator associated with the Levi-Civita connection of $g$.
\end{lemma}
This lemma and the following corollary are essential to prove uniqueness in theorem \ref{thm:distinitialvalueproblem}.
\begin{corollary}[Fresnel-Kirchhoff integral, cp. \cite{DimockAlgebrasOfLocal}]
\label{cor:normhypgreensidentity}
Assume that $P$ is also formally self-adjoint. If $u\in\cE(M,E^{*})$ is a solution to $Pu=0$, we have:
\begin{align}
\label{eq:uniquenessidentity}
\forall f\in\cD(M,E):\ \int_{M}(u,f)dV_{g} & = -\int_{\Sigma}\left((\nabla_{n}u,G(f))-(u,\nabla_{n}G(f))\right)dA_{g}.
\end{align}
\begin{proof}
We integrate \eqref{eq:normhypgreensidentity} with $f$ replaced by $G^{\mp}(f)$ in the domains $J_{\pm}(\Sigma)$ with the common boundary $\partial J_{\pm}(\Sigma)$, and apply Gauss' theorem:
\begin{align}
\label{eq:greenandgauss}
\int_{J_{\pm}(\Sigma)}(u,f) & = \mp\int_{\Sigma}\left((\nabla_{n}u,G^{\mp}(f))-(u,\nabla_{n}G^{\mp}(f))\right)dA_{g}.
\end{align}
Adding the two expression gives the result. \qed
\end{proof}
\end{corollary}
Clearly, the formulas for the solution \eqref{eq:solutionfromdata} and \eqref{eq:distsolutionfromdata} mimic \eqref{eq:uniquenessidentity}.
\begin{lemma}
\label{lem:restrictioncontinuity}
The maps $\iota^{*},\nu^{*}:\cE_{\usc}(M,E)\rightarrow\cD(\Sigma,E_{\Sigma})$ (see \eqref{eq:restrictionmaps}) are sequentially continous.
\begin{proof}
For $f\in\cE_{\usc}(M,E)$, take a converging sequence $\{f_{j}\}_{j=1}^{\infty}\subset\cE_{\usc}(M,E)$, i.e. there exists a compact subset $K\sqsubset M$, s.t.
\begin{align}
\label{eq:spacelikeconvergence}
& \forall j:\ \supp(f),\supp(f_{j})\subset J(K), \\[0.25cm]
& \forall k\in\N_{0}, K'\sqsubset M\ \textup{cpt.}:\ \lim_{j\rightarrow\infty}||f-f_{j}||_{C^{k}(K,E)}=0,
\end{align}
where $||f'||_{C^{k}(K',E)}:=\max_{n=1,...,k}\sup_{x\in K'}||\nabla^{n}f'(x)||_{g,g_{E}}$ for $f'\in\cE(M,E)$\footnote{See appendix \ref{sec:distmanifold} for the construction of the norms}. \\
We need to show that $\lim_{j\rightarrow\infty}\iota^{*}f_{j}=\iota^{*}f$ and $\lim_{j\rightarrow\infty}\nu^{*}f_{j}=\nu^{*}f$ in $\cD(\Sigma,E_{\Sigma})$. Because $M$ is globally hyperbolic and $\Sigma$ is a Cauchy surface, we know that $J(K)\cap\Sigma=:K''$ is compact. Moreover by the definition of the maps in question, we have
\begin{align}
\label{eq:restrictionsupp}
\supp(\iota^{*}f), \supp(\iota^{*}f_{j}), \supp(\nu^{*}f), \supp(\nu^{*}f_{j})\subset K'',
\end{align}
and
\begin{align}
\label{eq:restrictionconvergence}
||\iota^{*}f-\iota^{*}f_{j}||_{C^{k}(K'',E_{\Sigma})} & \leq ||f-f_{j}||_{C^{k}(K''',E)} \\[0.25cm]
||\nu^{*}f-\nu^{*}f_{j}||_{C^{k}(K'',E_{\Sigma})} & \leq ||f-f_{j}||_{C^{k}(K''',E)} 
\end{align}
for some compact subset $K'''\sqsubset M$, s.t. $K''\subset K'''$, and all $k\in\N_{0}$. This proves the statement. \qed
\end{proof}
\end{lemma}
\begin{proposition}
\label{prop:solutionmapcontinuity}
The compositions of adjoint maps,
\begin{align}
\label{eq:adjointcomp}
G'\circ(\iota^{*})',G'\circ(\nu^{*})':\cD'(\Sigma,E_{\Sigma}^{*})\rightarrow\cD'_{\Char P}(M,E^{*}),
\end{align}
are sequentially continuous w.r.t. H\"ormander's (pseudo-)topology on $\cD'_{\Char P}(M,E^{*})$ (see definition \ref{def:restrictedwavefrontsetdist} \& \cite{HoermanderFourierIntegralOperators, DabrowskiFunctionalPropertiesOf}).
\begin{proof}
For $u\in\cD'(\Sigma,E_{\Sigma}^{*})$, take a converging sequence $\{u_{j}\}_{j=1}^{\infty}\subset\cD'(\Sigma,E_{\Sigma}^{*})$, i.e.
\begin{align}
\label{eq:weakconvergencecauchy}
\forall f\in\cD(\Sigma,E^{*}):\ \lim_{j\rightarrow\infty}(u_{j},f) & = (u,f).
\end{align}
To show that $\lim_{j\rightarrow\infty}(G'\circ(\iota^{*})')(u_{j})=(G'\circ(\iota^{*})')(u)$ and $\lim_{j\rightarrow\infty}(G'\circ(\nu^{*})')(u_{j})=(G'\circ(\nu^{*})')(u)$, we use a characterization of convergence in $\cD'_{\Gamma}(M,E^{*}),\ \Gamma\subset T^{*}M\setminus\{0\}$ closed and conical, proven in \cite{DabrowskiFunctionalPropertiesOf}:\\[0.1cm]
Given a sequence $\{u_{j}\}_{j=1}^{\infty}\subset\cD'_{\Gamma}(M,E^{*})$, s.t. $\lim_{j\rightarrow\infty}(u_{j},v)=\lambda_{v}\in\C$ exists for all $v\in\cE'_{\Lambda}(M,E)$, then $\lim_{j\rightarrow\infty}u_{j}=u\in\cD'_{\Gamma}(M,E^{*})$ exists, s.t. $(u,v)=\lambda_{v}$ for all $v\in\cE'_{\Lambda}(M,E)$.\\[0.1cm]
Here, $\Lambda\subset T^{*}M\setminus\{0\} $ is the complement of the inversion of $\Gamma$:
\begin{align}
\label{eq:invertedconecomplement}
\Lambda := (\Gamma')^{c}=\{(x,k)\in T^{*}M\setminus\{0\}\ |\ (x,-k)\notin\Gamma\}.
\end{align}
Next, we observe that we have an extension $G:\cE'_{\Lambda}(M,E)\rightarrow\cE_{\usc}(M,E)$ in the sense of theorem \ref{thm:kernelextension}. To achieve this, we use the fact that Schwartz' kernel theorem gives us a distribution $K_{G}\in\cD'(M\times M,E\boxtimes E^{*})$, and check that the composition $K_{G}\circ v$ for $v\in\cE'_{\Lambda}(M,E)$ is well-defined. The wave front set of $K_{G}$ is well-known (cf. \cite{DuistermaatFourierIntegralOperators, RadzikowskiMicrolocalApproachTo}):
\begin{align}
\label{eq:propagatorwavefrontset}
\WF(K_{G}) & = \{(x,k;x',k')\in(\Char P)^{\times 2}\ |\ (x,k)\sim_{H_{P}}(x',-k')\},
\end{align}
where $(x,k)\sim_{H_{P}}(x',k')$ means that $(x,k),(x',k')\in T^{*}M$ lie on the same integral curve of the Hamiltonian vector field $H_{P}$ of the principal symbol of $P$\footnote{At coinciding point $x=x'$, we have $k=k'\neq0$.}. Because $P$ is normally hyperbolic, its principal symbol is given by the (inverse) metric $\sigma_{P}(x,k)=g^{ij}(x)k_{i}k_{j}$.\\
Thus, $\Char P=C^{*}M\setminus\{0\}$ is the co-light cone bundle without the zero section, and an integral curve of $H_{P}$ joining $(x,k)$ and $(x',k')$ is a null geodesic strip in $T^{*}M$, which projects to the null geodesic in $M$ from $x$ to $x'$ with co-tangents $k\in T^{*}_{x}M$ and $k'\in T^{*}_{x'}M$ (cf. \cite{RadzikowskiMicrolocalApproachTo}). Clearly, the Hamiltonian flow in $T^{*}M$ is in one-to-one correspondence with the null geodesics flow in $TM$ via the metric $g$. To apply theorem \ref{thm:kernelextension}, we need to check that
\begin{align}
\label{eq:propagatorcompositionwavefroncheck}
\WF(v)\cap(-\WF(K_{G})_{M_{2}|M_{1}}) & = \emptyset.
\end{align}
This is trivially satisfied, because $-\WF(K_{G})_{M_{2}|M_{1}}=\{(x',k')\in T^{*}M\setminus\{0\}\ |\ (x,0;x',k')\in \WF(K_{G})\}=\emptyset$ by \eqref{eq:propagatorwavefrontset}. Theorem \ref{thm:kernelextension} gives us information on the wave front set of $G(v)$, as well:
\begin{align}
\label{eq:propagatorcompositionwavefrontset}
\WF(G(v))\subset\underbrace{\WF(K_{G})_{M_{1}|M_{2}}}_{=\emptyset}\cup\underbrace{\WF'(K_{G})\circ \WF(v)}_{=\emptyset}=\emptyset,
\end{align}
by the definition of $\Lambda$ and \eqref{eq:propagatorwavefrontset}. It follows that $G(v)\in\cE(M,E)$. What remains to be checked, is that $G(v)\in\cE_{\usc}(M,E)$. To see this, we notice that $\supp(v)\sqsubset M$ is compact, because $v\in\cE'_{\Lambda}(M,E)$, which implies:
\begin{align}
\label{eq:propagatorcompositionsupport}
(G(v),f) & = -(v,G(f)) \\ \nonumber
 & = 0
\end{align}
for all $f\in\cD(M,E^{*})$, s.t. $\supp(f)\sqsubset(J(\supp(v)))^{c}$. Putting everything together, we find:
\begin{align}
\label{eq:hoermanderconvergence}
\lim_{j\rightarrow\infty}((G'\circ(\iota^{*})')(u_{j}),v) & = \lim_{j\rightarrow\infty}-((\iota^{*})'(u_{j}),\underbrace{G(v)}_{\in\cE_{\usc}(M,E)}) = \lim_{j\rightarrow\infty}-(u_{j},\underbrace{\iota^{*}G(v)}_{\in\cD(\Sigma,E^{*}_{\Sigma})}) \\ \nonumber
 & = -(u,\iota^{*}G(v)) = ((G'\circ(\iota^{*})')(u),v)\ \ \ \ \forall v\in\cE'_{\Lambda}(M,E).
\end{align}
The argument for $G'\circ(\nu^{*})'$ is analogous. \qed
\end{proof}
\end{proposition}
Now, we are in the position to prove theorem \ref{thm:distinitialvalueproblem}.
\begin{proof}
\begin{itemize}
	\item[\textbf{1.}] \textit{Existence}:\\[0.1cm]
Given $u_{0},u_{1}\in\cD'(\Sigma,E_{\Sigma}^{*})$, we use equation \eqref{eq:distsolutionfromdatawavefrontset} to define a solution $u\in\cD'_{\Char P}(M,E^{*})$:
\begin{align}
\label{eq:distsolutionansatz}
u & :=(G'\circ(\iota^{*})')(u_{1})-(G'\circ(\nu^{*})')(u_{0}).
\end{align}
We need to show that this solution satisfies $\iota^{*}u=u_{0},\ \nu^{*}u=u_{1}$. To this end, we observe that the extended maps
\begin{align}
\label{eq:distrestrictions}
\iota^{*},\nu^{*}:\cD'_{\Char P}(M,E^{*})\rightarrow\cD'(\Sigma,E_{\Sigma}^{*})
\end{align}
are well-defined and (sequentially) continuous by virtue of theorem \ref{thm:distpullback}:\\
The co-normal $N_{\iota}$ of $\iota:\Sigma\hookrightarrow M$ has empty intersection with the co-light cone bundle $\Char P=C^{*}M\setminus\{0\}$:
\begin{align}
\label{eq:emptyconormalintersection}
N_{\iota}\cap\Char P & = \emptyset.
\end{align}
Since $\cD(\Sigma,E_{\Sigma}^{*})$ is (sequentially) dense in $\cD'(\Sigma,E_{\Sigma}^{*})$, we find sequences $\{u_{0,j}\},\{u_{1,j}\}\subset\cD(M,E^{*})$, s.t. $\lim_{j\rightarrow\infty}u_{0,j}=u_{0}$ and $\lim_{j\rightarrow\infty}u_{1,j}=u_{1}$. Using the continuity of the maps \eqref{eq:distrestrictions} and proposition \ref{prop:solutionmapcontinuity}, we may write:
\begin{align}
\label{eq:compatibledistrestriction}
\iota^{*}u & = \iota^{*}(G'\circ(\iota^{*})')(u_{1})-\iota^{*}(G'\circ(\nu^{*})')(u_{0}) \\ \nonumber
 & = \iota^{*}(G'\circ(\iota^{*})')(\lim_{j\rightarrow\infty}u_{1,j})-\iota^{*}(G'\circ(\nu^{*})')(\lim_{j\rightarrow\infty}u_{0,j}) \\ \nonumber
 & = \lim_{j\rightarrow\infty}(\iota^{*}(G'\circ(\iota^{*})')(u_{1,j})-\iota^{*}(G'\circ(\nu^{*})')(u_{0,j})) \\ \nonumber
 & = \lim_{j\rightarrow\infty}u_{0,j} \\ \nonumber
 & = u_{0},
\end{align}
where we used the identities \eqref{eq:initialdatasymplecticform} after the next-to-last line.
The argument for $\nu^{*}u=u_{1}$ is analogous.
	\item[\textbf{2.}] \textit{Uniqueness}:\\[0.1cm]
If we want prove uniqueness of the solution \eqref{eq:distsolutionansatz} among the proper solutions of $Pu=0$, we first need to check that $u$ is indeed proper, and second, that any other proper solution $u'\in\cD'_{\Char P}(M,E^{*})$ with $\iota^{*}u'=u_{0},\ \nu^{*}u=u_{1}$ is identical to $u$, i.e. $u'\equiv u$.\\[0.1cm]
To see that $u$ is proper, we choose sequences $\{u_{0,j}\},\{u_{1,j}\}\subset\cD(M,E^{*})$ as before. Then, we set
\begin{align}
\label{eq:smoothsolutionsequence}
\forall j:\ u_{j} & :=(G'\circ(\iota^{*})')(u_{1,j})-(G'\circ(\nu^{*})')(u_{0,j})\in\cE_{\usc}(M,E^{*}),
\end{align}
which is a sequence of smooth solutions, s.t. $\lim_{j\rightarrow\infty}u_{j}=u$ in $\cD'_{\Char P}(M,E^{*})$, by proposition \ref{prop:solutionmapcontinuity}, \eqref{eq:solutionfromdata} and \eqref{eq:restrictionsolutionfromdata}.\\[0.1cm]
For the second statement, we observe that another solution $u'\neq u$ would imply the existence of a non-trivial, proper solution with vanishing initial data, i.e.
\begin{align}
\label{eq:zerodatasolution}
0\neq u'':=u-u'\in\cD'_{\Char P}(M,E^{*}):\ Pu''=0,\ \iota^{*}u''=0,\ \nu^{*}u''=0.
\end{align}
Thus, to conclude uniqueness, we need to show that the only proper solution with vanishing initial data is $u''\equiv 0$.
This can be done by an appeal to corollary \ref{cor:normhypgreensidentity}: Assume we are given a proper solution $u''$ as in \eqref{eq:zerodatasolution}. Then, we choose a approximating sequence $\{u''_{j}\}_{j=1}^{\infty}\subset\cE(M,E^{*})$, $\lim_{j\rightarrow\infty}u''_{j}=u'',\ \forall j:\ Pu''_{j}=0,$ and compute:
\begin{align}
\label{eq:greensidentityforuniqueness}
\forall f\in\cD(M,E):\ (u'',f) & = \lim_{j\rightarrow\infty}(u''_{j},f) \\ \nonumber
 & \stackrel{\textup{Cor.}\ref{cor:normhypgreensidentity}}{=} -\lim_{j\rightarrow\infty}((\nu^{*}u''_{j},\iota^{*}G(f))-(\iota^{*}u''_{j},\nu^{*}G(f))) \\ \nonumber
 & = 0,
\end{align}
where we use the continuity of $\iota^{*},\nu^{*}$ in the last line. But this contradicts $u''\neq0$.
	\item[\textbf{3.}] \textit{Continuous dependence on initial data}:\\[0.1cm]
	This is precisely the content of proposition \ref{prop:solutionmapcontinuity}. \qed
\end{itemize}
\end{proof}
\begin{remark}
\label{rem:linearmanifoldpropersolutions}
Clearly, the statement of theorem \ref{thm:distinitialvalueproblem} can be improved, if $M$ is a linear manifold and $P$ has constant coefficients, e.g. Minkowski space $\M$ and $P$ is the d'Alembertian. Namely, every distributional solution $u\in\cD'_{\Char P}(M,E^{*})$ is then a proper solution by virtue of the existence of an approximate identity $\{\phi_{\varepsilon}\}\subset\cD(M)$, $\lim_{\varepsilon\rightarrow0}\phi_{\varepsilon}=\delta_{0}$, H\"ormander's density theorem (see \cite{HoermanderTheAnalysisOf1}, p.262-263) and the convolution identities:
\begin{align}
\label{eq:convolutionidentities}
& \phi_{\varepsilon}\ast u\in\cE(M,E^{*}),\ \lim_{\varepsilon\rightarrow0}\phi_{\varepsilon}\ast u = u\ \textup{in}\ \cD'_{\Char P}(M,E^{*}), \\ \nonumber
& P(\phi_{\varepsilon}\ast u) = \phi_{\varepsilon}\ast (Pu) = 0.
\end{align}
Interestingly, \eqref{eq:advancedretarded} and \eqref{eq:propagatoridentity} tell us that $K_{G}$ is a proper, distributional (bi-)solution.
\end{remark}

\subsection{Quasifree states and the microlocal spectrum/Hadamard condition}
\label{sec:microlocalspectrumcondition}
We are now ready to turn our attention to the quantum theory associated with the classical setup of the previous subsection. From the exact sequence \eqref{eq:shortexact}, the well-posedness of the Cauchy problem for $P$ with initial data in $\cD(\Sigma,E_{\Sigma})$ and the identities \eqref{eq:initialdatasymplecticform}, we know, that we have a pair of isomorphic linear, symplectic spaces representing the space of (smooth) solution $\Sol^{\infty}_{0}(P)$ with compactly supported (smooth) initial data
\begin{align}
\label{eq:symplecticspaces}
\cD(M,E)/\im P_{|\cD(M,E)} & \cong\cD(\Sigma,E_{\Sigma})^{\oplus 2}.
\end{align}
The symplectic structures are given by (cf. \cite{BaerWaveEquationsOn})
\begin{align}
\label{eq:symplecticstructure}
\sigma^{M}([f],[f']) & =\int_{M}g_{E}(G(f),f')dV_{g} \\[0.25cm] \nonumber 
  &\!\!\!\!\!\stackrel{\textup{Cor.}\ref{cor:normhypgreensidentity}}{=}-\int_{\Sigma}\left(g_{E}(\nabla_{n}G(f),G(f'))-g_{E}(G(f),\nabla_{n}G(f'))\right)dA_{g} \\
  & =\sigma^{\Sigma}((f_{0},f_{1}),(f'_{0},f'_{1})),
\end{align}
for $[f],[f']\in\cD(M,E)/\im P_{|\cD(M,E)}$ and $(f_{0},f_{1}),(f'_{0},f'_{1})\in\cD(\Sigma,E_{\Sigma})^{\oplus 2}$, which are identified by virtue of the isomorphism \eqref{eq:symplecticspaces}. The expressions are well-defined, because $\forall f\in\cD(M,E):G(Pf)=0$ and $(\nabla_{n}G(f))_{|\Sigma},G(f)_{|\Sigma}$ defines the initial data for the solution $G(f)\in\Sol^{\infty}_{0}(P)$.\\[0.1cm]
This allows us to consider the space $\Sol^{\infty}_{0}(P)$ as a symplectic space, with symplectic structure $\sigma$, and it is well-known that we can associate a ($C^{*}$-)Weyl algebra $\cW_{P}$ with it\footnote{See \cite{BaerWaveEquationsOn} for a detailed exposition with an emphasis on local covariance \cite{BrunettiTheGenerallyCovariant} and functoriality of the construction. The are alternative algebraic structures, as well, e.g. the Resolvent algebra \cite{BuchholzTheResolventAlgebra1, BuchholzTheResolventAlgebra2}. These could be worthwhile to consider, since the Weyl algebra only admits a very restricted set of dynamics ($C^{*}$-automorphism 1-parameter groups).}. This algebra is generated by the Weyl elements $W(G(f)),\ G(f)\in\Sol^{\infty}_{0}(P),$ subject to the CCR relations in Weyl form:
\begin{align}
\label{eq:weylrelations}
W(G(f))W(G(f')) & = e^{-\frac{i}{2}\sigma(G(f),G(f'))}W(G(f+f')).
\end{align}
A Hilbert space representation of the quantum system defined by the Weyl algebra $\cW_{P}$ is obtained by specifying an (algebraic) state $\omega:\cW_{P}\rightarrow\C$ and passing to the GNS representation $(\mathfrak{H}_{\omega},\pi_{\omega},\Omega_{\omega})$ (see \cite{BratteliOperatorAlgebrasAnd1,BratteliOperatorAlgebrasAnd2} for a detailed account on the algebraic formulation of quantum theory). An important class of states on $\cW_{P}$ is given by the (regular) quasi-free states, i.e. states $\omega$, which are solely determined via their two-point function(al) (cf. \cite{BrunettiTheGenerallyCovariant}):
\begin{align}
\label{eq:quasifreestate}
\omega(W(G(f))) & = e^{-\frac{1}{2}\omega_{2}(f,f)}, \\[0.25cm] \nonumber
\omega_{2}(f,f') & =-\frac{\partial^{2}}{\partial t\partial s}_{|t,s=0}\omega(W(G(tf))W(G(sf'))).
\end{align}
It is important for the following that this definition requires $\omega_{2}:\cD(M,E)^{\times 2}\rightarrow\C$ to be a distributional (bi-)solution for $P$, i.e. $\forall f,f'\in\cD(M,E):\ \omega_{2}(Pf,f)=0=\omega_{2}(f,Pf')$. Among the quasi-free states are the physically important Hadamard states, which can be regarded as a replacement for the vacuum state of quantum field theory on Minkowski space, since they can be characterized as having a short-distance singularity structure analogous to that of the Minkowski vacuum (cf. \cite{WaldQuantumFieldTheory}, and \cite{BrunettiTheGenerallyCovariant} for important structural properties of the folium of Hadamard states). In a seminal paper \cite{RadzikowskiMicrolocalApproachTo}, Radzikowski showed that Hadamard states are equivalently characterized by a specific form of the wave front set of their two-point function(al) (cp. \eqref{eq:propagatorwavefrontset}):
\begin{align}
\label{eq:microlocalspectrumcondition}
\WF(\omega_{2}) & = \WF(K_{G})\cap(C_{+}^{*}M\times C^{*}_{-}M) \\ \nonumber
 & = \{(x,k;x',k')\in(\Char P)^{\times 2}\ |\ (x,k)\sim_{H_{P}}(x',-k'), k\ \textup{is\ future-directed}\},
\end{align}
where $C^{*}_{\pm}M$ are the future-/past-directed, co-light cone bundles of $M$\footnote{Along the diagonal $\Delta_{M}\subset M\times M$ we have $\WF(\omega_{2})_{|\Delta_{M}}=\{(x,k;x-k)\in(T^{*}M)^{\times2}\setminus\{0\}\ |\ k\in C^{*}_{+|x}M\setminus\{0\}\}$.}. Thus, the two-point function(al) of a Hadamard state has a wave front set resembling the spectral condition, i.e. positivity of the energy, of quantum field theory on Minkowski space in a microlocal fashion, which justifies the name \textit{microlocal spectrum condition} for \eqref{eq:microlocalspectrumcondition}. What is even more remarkable, is the fact that the microlocal spectrum condition admits a generalization to allow for locally covariant treatment of interacting quantum field on curved spacetimes in a perturbative setting \cite{BrunettiTheMicrolocalSpectrum, BrunettiMicrolocalAnalysisAnd, HollandsExistenceOfLocal, KratzertSingularityStructureOf, HollandsTheHadamardCondition, DAntoniNuclearityLocalQuasiequivalence, DappiaggiTheExtendedAlgebra, SandersTheLocallyCovariant, RejznerFermionicFieldsIn, HollandsRenormalizedQuantumYangMills, FredenhagenBatalinVilkoviskyFormalism1, FredenhagenBatalinVilkoviskyFormalism2, FredenhagenLocalCovarianceAnd}. A crucial observation in this respect is the fact, that a Hadamard state defines a \textit{Feynman propagator} \cite{RadzikowskiMicrolocalApproachTo}:
\begin{align}
\label{eq:feynmannpropagator}
\omega_{F}:=i\omega_{2}-K_{G^{-}},
\end{align}
such that it agrees with the state-independent Feynman parametrix $K_{G_{F}}$ of Duistermaat and H\"ormander \cite{DuistermaatFourierIntegralOperators}. Since we are interested in the application of this condition in quantum theories, which are based on an initial value or Hamiltonian formulation, we need a to transfer \eqref{eq:microlocalspectrumcondition} to such a framework. One possibility to achieve this, suggests itself by the observation that $\omega_{2}\in\cD'_{\Char P}(M\times M,E^{*}\boxtimes E^{*})$ is a distributional (bi-)solution for $P$. If we additionally assume that $\omega_{2}$ is proper in the sense of theorem \ref{thm:distinitialvalueproblem}, we may associate unique initial data in $\cD'(\Sigma\times\Sigma,E^{*}_{\Sigma}\boxtimes E^{*}_{\Sigma})$ with it:
\begin{align}
\label{eq:twopointinitialdata}
\omega_{2,00}:=(\iota^{*}\times\iota^{*})\omega_{2},\ & \ \omega_{2,01}:=(\iota^{*}\times\nu^{*})\omega_{2}, \\[0.25cm] \nonumber
\omega_{2,10}:=(\nu^{*}\times\iota^{*})\omega_{2},\ &\ \omega_{2,00}:=(\nu^{*}\times\nu^{*})\omega_{2},
\end{align}
\begin{align}
\label{eq:twopointsolutionfromdata}
\omega_{2} = &\hspace{0.45cm}((G'\circ(\nu^{*})')\times(G'\circ(\nu^{*})'))\omega_{2,00}-((G'\circ(\nu^{*})')\times(G'\circ(\iota^{*})'))\omega_{2,01} \\[0.25cm] \nonumber
 & -((G'\circ(\iota^{*})')\times(G'\circ(\nu^{*})'))\omega_{2,10}+((G'\circ(\iota^{*})')\times(G'\circ(\iota^{*})'))\omega_{2,11}.
\end{align}
Equation \eqref{eq:twopointsolutionfromdata} tells us that the data \eqref{eq:twopointinitialdata} is exactly what we need to evaluate \eqref{eq:quasifreestate}, if we use the initial data description for the solution space $\Sol^{\infty}_{0}(P)\cong\cD(\Sigma,E_{\Sigma})^{\oplus 2}$ (see \eqref{eq:symplecticspaces}):
\begin{align}
\label{eq:twopointevaluation}
\omega_{2}(f,f') & = \omega_{2,00}(f_{1},f'_{1}) - \omega_{2,01}(f_{1},f'_{0}) - \omega_{2,10}(f_{0},f'_{1}) + \omega_{2,11}(f_{0},f'_{0}).
\end{align}
This motivates our main theorem:
\begin{theorem}[The microlocal spectrum of Hadamard initial data]
\label{thm:hadamardinitialdata}
Given a quasi-free Hadamard state $\omega$. If the two-point function(al) $\omega_{2}\in\cD'_{\Char P}(M\times M,E^{*}\boxtimes E^{*})$ is proper (in the sense of theorem \ref{thm:distinitialvalueproblem}), the initial data \eqref{eq:twopointinitialdata} satisfies the bound
\begin{align}
\label{eq:twopointinitialdatawavefrontset}
\bigcup_{i,j=0}^{1}\WF(\omega_{2,ij}) & =N_{\Delta}\setminus\{0\},
\end{align}
where $N_{\Delta}=\{(\Delta(x),(k,k'))\in T^{*}\Sigma^{\times2}\ |\ k=-k'\}$ is the co-normal of the diagonal map $\Delta:\Sigma\rightarrow\Sigma\times\Sigma$.
\end{theorem}
Before we start the proof, let us outline the rough idea and why \eqref{eq:twopointinitialdatawavefrontset} is plausible from the point of view of canonical quantization on $\Sigma$, we will follow:\\[0.1cm]
The wave front set of an initial datum $u_{|\Sigma}\in\cD'(\Sigma,E^{*}_{\Sigma})$ for a (proper) solution $u\in\cD'(M,E^{*})$ can be estimated by the tools of microlocal analysis from $\WF(u)$, because, on the one hand, $u$ arises from $u_{|\Sigma}$ by composition with the causal propagator $K_{G}$ of $P$ (see \eqref{eq:twopointsolutionfromdata}), and on the other hand, $u_{|\Sigma}$ is the restriction of $u$. Thus, the knowledge of $\WF(\omega_{2})$ gives us a two-sided estimate on the wave front sets $\WF(\omega_{2,ij}),\ i,j=0,1$. Furthermore, since the $K_{G}$ propagates singularities along the co-light cone bundle (see theorem \ref{thm:characteristicwavefrontset}), the initial data for a Hadamard state must contain enough singular directions to satisfy microlocal spectrum condition, which is the reason for \eqref{eq:twopointinitialdatawavefrontset}. In view of canonical quantization, where
\begin{align}
\label{eq:ccr}
\textup{``}[\Phi_{|\Sigma}(x),(\nabla_{n}\Phi)_{|\Sigma}(x')] & \sim \delta^{m-1}(x,x')\textup{''},
\end{align}
and \eqref{eq:spatialdeltawavefrontset} below, this seems adequate.
\begin{proof}
We prove the theorem by showing the inclusions $\bigcup_{i,j=0}^{1}\WF(\omega_{2,ij})\subset N_{\Delta}\setminus\{0\}$ and $\bigcup_{i,j=0}^{1}\WF(\omega_{2,ij})\supset N_{\Delta}\setminus\{0\}$.
\begin{itemize}
	\item[\textbf{1.}] $\bigcup_{i,j=0}^{1}\WF(\omega_{2,ij})\subset N_{\Delta}\setminus\{0\}$:\\[0.1cm]
We have $\WF(\nu^{*}u)\subset\iota^{*}\WF(u)$ for any $u\in\cD'_{\Char P}(M,E^{*})$, because $\WF(Pu)\subset\WF(u)$ for any differential operator (see corollary \ref{cor:differentialwavefrontset}). Thus, if we show $\WF(\omega_{2,00})\subset N_{\Delta}\setminus\{0\}$, the first inclusion follows. Using theorem \ref{thm:distpullback}, we find:
\begin{align}
\label{eq:wavefrontsetinclusion1}
\WF(\omega_{2,00}) & \subset(\iota^{*}\times\iota^{*})\WF(\omega_{2}) \\ \nonumber
 &\!\!\!\! \stackrel{\eqref{eq:microlocalspectrumcondition}}{=}\!\!\{(x,d\iota_{|x}^{*}k;x',d\iota_{|x'}^{*}k')\in(T^{*}\Sigma)^{\times2}\setminus\{0\} | (\iota(x),k)\sim_{H_{P}}(\iota(x'),-k'), k\in C^{*}_{+|\iota(x)}M\} \\ \nonumber
 & =\{(x,\kappa;x,-\kappa)\in(T^{*}\Sigma\setminus\{0\})^{\times2}\}=N_{\Delta}\setminus\{0\}.
\end{align}
The last line follows, because:
\begin{itemize}
	\item[(a)] $(\iota(x),k)\sim_{H_{P}}(\iota(x'),-k')$ requires $\iota(x),\iota(x')\in\Sigma$ to lie on a common null geodesic or be equal. Thus, the only possibility is $(\iota(x'),k')=(\iota(x),-k)$, since $\Sigma$ is acausal (cf. \cite{ONeillSemiRiemannianGeometry}).
	\item[(b)] $d\iota_{|x}^{*}:\Char P_{|\iota(x)}\cap C^{*}_{\pm|\iota(x)}M\rightarrow T^{*}_{x}\Sigma$ is an isomorphism of conical sets for every $x\in\Sigma$.
 \end{itemize}
	\item[\textbf{2.}] $\bigcup_{i,j=0}^{1}\WF(\omega_{2,ij})\supset N_{\Delta}\setminus\{0\}$:\\[0.1cm]
From equation \eqref{eq:twopointsolutionfromdata}, we find:
\begin{align}
\label{eq:wavefrontsetinclusion2}
\WF(\omega_{2}) & \subset\WF(((G'\circ(\nu^{*})')\times(G'\circ(\nu^{*})'))\omega_{2,00})\cup\WF(((G'\circ(\nu^{*})')\times(G'\circ(\iota^{*})'))\omega_{2,01}) \\[0.25cm] \nonumber
&\ \ \ \cup\WF(((G'\circ(\iota^{*})')\times(G'\circ(\nu^{*})'))\omega_{2,10})\cup\WF(((G'\circ(\iota^{*})')\times(G'\circ(\iota^{*})'))\omega_{2,00}).
\end{align}
Thus, we may derive the second inclusion, if we compute a bound on the wave front set of the individual contributions in \eqref{eq:twopointsolutionfromdata}. This can be done with the help of theorem \ref{thm:kernelextension}, because
\begin{align}
\label{eq:restrictedpropagatorcomposition}
((G'\circ(\nu^{*})')\times(G'\circ(\nu^{*})'))\omega_{2,00} & = \omega_{2,00}\circ((\iota^{*}\circ G)\times(\iota^{*}\circ G)) \\[0.25cm] \nonumber
 & = \omega_{2,00}\circ((\iota^{*}\times\id^{*}_{M})K_{G})^{\otimes2}\ \textup{etc}.,
\end{align}
where the last line is interpreted as composition of distribution
\begin{align}
\label{eq:twopointcomposition}
\xymatrix{
\cD'(\Sigma^{\times2},(E^{*}_{\Sigma})^{\boxtimes2})\times\cD'(\Sigma^{\times2}\times M^{\times2},E_{\Sigma}^{\boxtimes2}\boxtimes(E^{*})^{\boxtimes2})\ar[r] & \cD'(M^{\times2},(E^{*})^{\boxtimes2}).
}
\end{align}
Furthermore, it is important that we have (see corollary \ref{cor:differentialwavefrontset} \& theorem \ref{thm:distpullback}):
\begin{align}
\label{eq:restricteddifferentialpropagatorwavefrontset}
\WF((\nu^{*}\times\id^{*}_{M})K_{G}) & \subset(\iota^{*}\times\id^{*}_{M})WF(K_{G}) \\ \nonumber
 &\!\!\!\!\stackrel{\eqref{eq:microlocalspectrumcondition}}{=}\!\!\{(x,(d\iota_{|x})^{*}k;x',k')\in(T^{*}\Sigma\setminus\{0\})\times\Char P | (\iota(x),k)\sim_{H_{P}}(x',-k')\}.
\end{align}
By virtue of proposition \ref{prop:tensorproductwavefrontset}, we can determine the wave front set of $((\iota^{*}\times\id^{*}_{M})K_{G})^{\otimes2}$:
\begin{align}
\label{eq:restrictedpropagatorwavefrontset}
\WF(((\iota^{*}\times\id^{*}_{M})K_{G})^{\otimes2}) & \subset\WF((\iota^{*}\times\id^{*}_{M})K_{G})^{\times2} \\[0.25cm] \nonumber
 &\ \ \cup(\supp((\iota^{*}\times\id^{*}_{M})K_{G})\times\{0\})\times\WF((\iota^{*}\times\id^{*}_{M})K_{G}) \\[0.25cm] \nonumber
 &\ \ \cup\WF((\iota^{*}\times\id^{*}_{M})K_{G})\times(\supp((\iota^{*}\times\id^{*}_{M})K_{G})\times\{0\}).
\end{align}
Theorem \ref{thm:kernelextension} tells us that the wave front set of \eqref{eq:restrictedpropagatorcomposition} obeys:
\begin{align}
\label{eq:compositionwavefrontset1}
\WF(((G'\circ(\nu^{*})')\times(G'\circ(\nu^{*})'))\omega_{2,00}) & \subset\WF(((\iota^{*}\times\id^{*}_{M})K_{G})^{\otimes2})_{M\times M|\Sigma\times\Sigma} \\[0.25cm] \nonumber
 &\ \ \cup\WF'(\omega_{2,00})\circ\WF(((\iota^{*}\times\id^{*}_{M})K_{G})^{\otimes2})\ \textup{etc}.
\end{align}
Putting \eqref{eq:restricteddifferentialpropagatorwavefrontset}, \eqref{eq:restrictedpropagatorwavefrontset} \& \eqref{eq:compositionwavefrontset1} together (see also \eqref{eq:projectedwavefrontset}), we find:
\begin{align}
\label{eq:compositionwavefrontset2}
\WF(((G'\circ(\nu^{*})')\times(G'\circ(\nu^{*})'))\omega_{2,00}) & \subset\WF'(\omega_{2,00})\circ\WF(((\iota^{*}\times\id^{*}_{M})K_{G})^{\otimes2})\ \textup{etc}.,
\end{align}
because
\begin{align}
\label{eq:compositionwavefrontset2a} \nonumber
\WF((\iota^{*}\times\id^{*}_{M})K_{G})_{M|\Sigma} & =\{(x',k')\in T^{*}M\setminus\{0\}\ |\ (x,0;x',k')\in\WF((\iota^{*}\times\id^{*}_{M})K_{G}),\ x\in\Sigma\} \\
 & =\emptyset.
\end{align}
Thus, what remains to be computed, is the composition of wave front sets $\WF'(\omega_{2,00})\circ\WF(((\iota^{*}\times\id^{*}_{M})K_{G})^{\otimes2})$, which can be done by means of \eqref{eq:restricteddifferentialpropagatorwavefrontset} (second line):
\begin{align}
\label{eq:compositionwavefrontset3}
 &\WF'(\omega_{2,00})\circ\WF(((\iota^{*}\times\id^{*}_{M})K_{G})^{\otimes2}) \\[0.25cm] \nonumber
 & = \{(x,k;x'k')\in(T^{*}M)^{\times 2}\setminus\{0\}\ |\ (x'',-\kappa;x''',-\kappa')\in\WF(\omega_{2,00}), \\ \nonumber
 &\ \ \ \ \ \ (x'',\kappa;x''',\kappa';x,k;x',k')\in\WF(((\iota^{*}\times\id^{*}_{M})K_{G})^{\otimes2})\} \\[0.25cm] \nonumber
 & \subset \{(x,k;x'k')\in(T^{*}M)^{\times 2}\setminus\{0\}\ |\ (x'',-(d\iota_{|x''})^{*}k'';x''',-(d\iota_{|x'''})^{*}k''')\in\WF(\omega_{2,00}), \\ \nonumber
 &\ \ \ \ \ \ (x'',(d\iota_{|x''})^{*}k'';x''',(d\iota_{|x'''})^{*}k''';x,k;x',k')\in(T^{*}\Sigma)^{\times2}\times(T^{*}M)^{\times2}\setminus\{0\},\\ \nonumber
 &\ \ \ \ \ \ (\iota(x''),x),(\iota(x'''),x')\in\supp(K_{G}),\ (\iota(x''),k'';x,k),(\iota(x'''),k''';x',k')\in(\Char P)^{\times2}\cup\{0\}, \\ \nonumber
 &\ \ \ \ \ \ (\iota(x''),k'')\sim_{H_{P}}(x,-k),(\iota(x'''),k''')\sim_{H_{P}}(x',-k')\ \textup{if}\ k\neq0, k'\neq0\}\ \textup{etc}.
\end{align}
If we combine the rather complicated looking expression in the last line with the requirement \eqref{eq:wavefrontsetinclusion2} and microlocal spectrum condition \eqref{eq:microlocalspectrumcondition}, we realize that we have to require that the wave front sets $\WF(\omega_{2,00})$ etc. contain elements $(x'',-(d\iota_{|x''})^{*}k'';x'',(d\iota_{|x''})^{*}k'')\in(T^{*}\Sigma)^{\times2}\setminus\{0\},\ k''\in C^{*}_{-|\iota(x'')}M$. This is the case, because the relations
\begin{align}
\label{eq:compositionrelations}
(x,k) & \sim_{H_{P}}(x',-k'),\ k\in C_{+|x}^{*}M, \\ \nonumber
(\iota(x''),-k'') & \sim_{H_{P}}(x,k) \\ \nonumber
(\iota(x'''),k''') & \sim_{H_{P}}(x',-k')
\end{align}
have to hold simultaneously, which can only be satisfied if $\iota(x'')=\iota(x''')$, implying $-k''=k'''$, since $\Sigma$ is acausal (cf. \cite{ONeillSemiRiemannianGeometry}). But, elements of the form $(x'',-(d\iota_{|x''})^{*}k'';x'',(d\iota_{|x''})^{*}k'')\in(T^{*}\Sigma)^{\times2}\setminus\{0\}$ are exactly those of the co-normal set $N_{\Delta}\setminus\{0\}$, and $d\iota_{|x}^{*}:\Char P_{|\iota(x)}\cap C^{*}_{\pm|\iota(x)}M\rightarrow T^{*}_{x}\Sigma$ is an isomorphism of conical sets for every $x\in\Sigma$. This implies the second inclusion. \qed
\end{itemize}
\end{proof}
To illustrate \eqref{eq:twopointinitialdatawavefrontset}, we consider the ground state of Klein-Gordon field of mass $m>0$ on an ultra-static spacetime, which is known to be Hadamard \cite{FullingSingularityStructureOf2}, and includes the important case of the vacuum state in Minkowski space.
\begin{example}
\label{ex:ultrastaticgroundstate}
Take an ultra-static spacetime $(M,g)=(\R,-dt^{2})\times(\Sigma,h)$, where $(\Sigma,h)$ is a complete, d-dimensional, Riemannian manifold, e.g. $\R^{3}$ with its standard metric. The analysis of the Klein-Gordon operator $\fbox{}_{g}+m^{2}=\partial^{2}_{t}-\Delta_{h}+m^{2}$ is conveniently phrased in terms of the strictly positive ($m>0$), elliptic operator $D=-\Delta_{h}+m^{2}$ (cf. \cite{FullingAspectsOfQuantum} for a detailed exposition), which is essentially self-adjoint together with all its natural powers on $L^{2}(\Sigma,dV_{h})$ with dense domain $\cD(\Sigma)$ \cite{ChernoffEssentialSelfAdjointness}. In an abuse of notation, we denote the closure of $D$ by the same letter. Then, the operator $\sqrt{D}$ is a strictly positive, elliptic, self-adjoint, pseudo-differential operator, and we have the important property \cite{JunkerHadamardStatesAdiabatic, HoermanderTheAnalysisOf3}
\begin{align}
\label{eq:ellipticwavefrontset}
\WF(\sqrt{D}u) & = \WF(u),\ u\in\cE'(\Sigma).
\end{align}
Moreover, $\sqrt{D}$ admits a suitable (Borel) functional calculus \cite{JunkerHadamardStatesAdiabatic}. Since $\cD(\Sigma)$ is nuclear, we can find a spectral resolution of the kernel $K_{D}\in\cD'(\Sigma\times\Sigma)$ of $D$ as an integral operator in $L^{2}(\Sigma,dV_{h})$ \cite{GelfandGeneralizedFunctions4, DunfordLinearOperators2}:
\begin{align}
\label{eq:nuclearspectralresolution}
K_{D} & = \int_{\sigma(\sqrt{D})}\omega^{2} f_{\omega}\bar{f}_{\omega}d\mu(\omega),
\end{align}
where $f_{\omega}\in\cE(\Sigma), Df_{\omega}=\omega^{2}f_{\omega},$ and $\sigma(K)\ni\omega\geq m>0$. This said, the two-point funtion(al) of the ground state for the quantum field theory of the Klein-Gordon field can be expressed as an integral operator in $L^{2}(\Sigma,dV_{h})$, as well:
\begin{align}
\label{eq:ultrastaticgroundstate}
\omega_{2,\infty}(t,x;t',x') & = \int_{\sigma(\sqrt{D})}\frac{e^{-i\omega(t-t')}}{2\omega}f_{\omega}(x)\bar{f}_{\omega}(x')d\mu(\omega),
\end{align}
which allows us to compute the initial data relative to $\Sigma_{0}=\{0\}\times\Sigma$:
\begin{align}
\label{eq:ultrastaticinitialdata}
\omega_{2,\infty,00}(x,x') & = \int_{\sigma(\sqrt{D})}\frac{1}{2\omega}f_{\omega}(x)\bar{f}_{\omega}(x')d\mu(\omega) = \frac{1}{2}\sqrt{D}^{-1}\frac{\delta^{(d)}(x,x')}{\sqrt{h}(x)}, \\[0.25cm] \nonumber
\omega_{2,\infty,01}(x,x')= -\omega_{2,\infty,10}(x,x') & = \int_{\sigma(\sqrt{D})}f_{\omega}(x)\bar{f}_{\omega}(x')d\mu(\omega) = -i\frac{\delta^{(d)}(x,x')}{2\sqrt{h}(x)}, \\[0.25cm] \nonumber
\omega_{2,\infty,11}(x,x') & = \int_{\sigma(\sqrt{D})}\frac{\omega}{2}f_{\omega}(x)\bar{f}_{\omega}(x')d\mu(\omega) = \frac{1}{2}\sqrt{D}\frac{\delta^{(d)}(x,x')}{\sqrt{h}(x)}.
\end{align}
This implies for the wave front sets
\begin{align}
\label{eq:ultrastaticwavefrontsets}
\WF(\omega_{2,\infty,00})=\WF(\omega_{2,\infty,01})=\WF(\omega_{2,\infty,10})=\WF(\omega_{2,\infty,11})=\WF(\sqrt{h}\delta^{(d)}),
\end{align}
because of \eqref{eq:ellipticwavefrontset} and the smoothness of $\sqrt{h}$. Furthermore, we have in a local coordinate system $U\subset\R^{d}$
\begin{align}
\label{eq:spatialdeltawavefrontset}
(\sqrt{h}\delta^{(d)})(f e^{-ik(\ .\ )},g e^{-ik'(\ .\ )})=\int_{U}f(x)g(x)e^{-i(k+k')x}\sqrt{h}(x)d^{d}x,\ f,g\in\cD(U),
\end{align}
and thus $\WF(\sqrt{h}\delta^{(d)})=N_{\Delta}\setminus\{0\}$. This shows that \eqref{eq:twopointinitialdatawavefrontset} holds and is maximally saturated.
\end{example}

\section{Concluding remarks}
\label{sec:remarks}
In the previous section, we have shown that the initial data of a Hadamard state with proper two-point function(al) must satisfy the bound
\begin{align}
\label{eq:hadamardbound}
\bigcup_{i,j=0}^{1}\WF(\omega_{2,ij}) & =N_{\Delta}\setminus\{0\},
\end{align}
where $N_{\Delta}=\{(\Delta(x),(k,k'))\in T^{*}\Sigma^{\times2}\ |\ k=-k'\}$ is the co-normal of the diagonal map $\Delta:\Sigma\rightarrow\Sigma\times\Sigma$. This bound on the wave front sets of the initial data could be regarded as optimal in the following sense: If the two-point function(al) $\omega_{2}$ has only a single non-smooth initial datum, e.g. $\WF(\omega_{2,00})\neq\emptyset$, the bound \eqref{eq:twopointinitialdatawavefrontset} is strict:
\begin{align}
\label{eq:optimalbound}
\omega_{2}\ \textup{satisfies\ the\ microlocal\ spectrum\ condition} & \Rightarrow \WF(\omega_{2,00})=N_{\Delta}\setminus\{0\}.
\end{align}
But, as we will see below (\eqref{eq:twopointevolution}) (cp. also \eqref{eq:twopointsolutionfromdata}), the requirement that only one initial datum is non-smooth is unstable w.r.t. the dynamics. Furthermore, the theorem tells us that the wave front sets of the initial data are already restricted in terms of the geometry of the Cauchy surface $\Sigma$, only. There is no reference to the metric (or causal) structure of $M$, besides the fact that $\Sigma$ is Cauchy. Although, it is true that assigning initial data to a solution depends on the spacetime metric via the maps \eqref{eq:restrictionmaps}, choosing initial data does not depend on this structure. \eqref{eq:optimalbound} is background independent in this sense. Thus, we have a condition that is applicable to settings, where no spacetime metric is available, e.g. loop quantum gravity.  \\
Interestingly, the proof of theorem \ref{thm:hadamardinitialdata} shows that the form of the (primed!) wave front set $\WF'(\omega_{2})$ required by the microlocal spectrum condition represents a minimal, conical, $H_{P}$-invariant (cf. theorem \ref{thm:characteristicwavefrontset}) subset of $(C^{*}M\setminus\{0\})^{\times2}$, s.t. the pullback of the restriction of $\WF(\omega_{2})$ (unprimed!)  to the diagonal $\Delta_{M}$ in $M\times M$ gives the full, non-zero co-normal of the diagonal $\Delta_{\Sigma}$ in $\Sigma\times\Sigma$, i.e. $(\iota\times\iota)^{*}\WF(\omega_{2})_{|\Delta_{M}}=N_{\Delta}\setminus\{0\}$\footnote{$-\WF(\omega_{2})$ satisfies these conditions, too, which would correspond to choosing the opposite time orientation on $M$.}. The minimality of $\WF'(\omega_{2})$ follows from the fact that a subset $V\subset(T^{*}M)^{\times2}\setminus\{0\}$ with these properties must contain $d\Delta_{|x}(C^{*}_{\pm|x}M\setminus\{0\}),\ x\in M,$ when restricted to the diagonal $\Delta_{M}$\footnote{We denote the diagonal map $M\rightarrow M\times M$ also by $\Delta$.}. \\
On the other hand, the proof also shows that initial data subject to \eqref{eq:hadamardbound} does not uniquely correspond to a (bi-)solution with a wave front set satisfying the microlocal spectrum condition. For example, the initial data of the recently proposed S-J vacuum \cite{AfshordiADistinguishedVacuum} for the Klein-Gordan field of mass $m>0$ on an ultra-static slab spacetime $(M,g)=(I_{\tau},-dt^{2})\times(\Sigma,h)$ satisfies, and even saturates, this bound, as well, but does not define a Hadamard state in general \cite{FewsterOnARecent, BrumVacuumLikeHadamard}. Here, $I_{\tau}=(-\tau,\tau),\ \tau>0$ and $(\Sigma,h)$ is a compact, d-dimensional, Riemannian manifold. To see that the initial data of the S-J vacuum respects \eqref{eq:optimalbound}, we argue in same way as in example \ref{ex:ultrastaticgroundstate}:\\[0.1cm]
Since $\Sigma$ is assumed to be compact, the spectral measure $\mu$ in \eqref{eq:nuclearspectralresolution} is supported in a countable set of points $\{\omega_{j}\}_{j\in J}$, and the two-point function(al) of the S-J vacuum is given by (cf. \cite{FewsterOnARecent}):
\begin{align}
\label{eq:SJtwopoint}
\omega_{2,\textup{S-J}}(t,x;t',x') & = \sum_{j\in J}\frac{1}{2\omega_{j}(1-\delta_{j})}(e^{-i\omega_{j}t}+i\delta_{j}\sin(\omega_{j}t))(e^{i\omega_{j}t'}-i\delta_{j}\sin(\omega_{j}t'))f_{\omega_{j}}(x)\bar{f}_{\omega_{j}}(x'),
\end{align}
where $1-\delta_{j}=\sqrt{\frac{1-\sinc(2\omega_{j}\tau)}{1+\sinc(2\omega_{j}\tau)}}, j\in J$. It is important for the following that $1-\delta_{j}$ is strictly bounded away from zero and bounded above as a function of $\omega_{j}$, because $\omega_{j}\geq m>0$. The initial data relative to $\Sigma_{0}=\{0\}\times\Sigma$ takes form:
\begin{align}
\label{eq:SJinitialdata}
\omega_{2,\textup{S-J},00}(x,x') & = \sum_{j\in J}\frac{1}{2\omega_{j}(1-\delta_{j})}f_{\omega_{j}}(x)\bar{f}_{\omega_{j}}(x') = \frac{1}{2}(\sqrt{D}(1-\delta)(\sqrt{D}))^{-1}\frac{\delta^{(d)}(x,x')}{\sqrt{h}(x)}, \\[0.25cm] \nonumber
\omega_{2,\textup{S-J},01}(x,x')= -\omega_{2,\textup{S-J},10} & = -\frac{i}{2}\sum_{j\in J}f_{\omega_{j}}(x)\bar{f}_{\omega_{j}}(x') = -i\frac{\delta^{(d)}(x,x')}{2\sqrt{h}(x)}, \\[0.25cm] \nonumber
\omega_{2,\textup{S-J},11}(x,x') & = \sum_{j\in J}\frac{\omega_{j}(1-\delta_{j})}{2} = \frac{1}{2}\sqrt{D}(1-\delta)(\sqrt{D})\frac{\delta^{(d)}(x,x')}{\sqrt{h}(x)}.
\end{align}
Here, we defined the elliptic, self-adjoint, pseudo-differential operator $(1-\delta)(\sqrt{D})$ by the functional calculus of $\sqrt{D}$ (see example \ref{ex:ultrastaticgroundstate}). By a similar argument as above, we have:
\begin{align}
\label{eq:SJinitaldatawavefrontset}
\WF(\omega_{2,\textup{S-J},00})=\WF(\omega_{2,\textup{S-J},01})=\WF(\omega_{2,\textup{S-J},10})=\WF(\omega_{2,\textup{S-J},11})=N_{\Delta}\setminus\{0\}.
\end{align}
Summarizing, we expect that \eqref{eq:hadamardbound} does not capture the full dynamical content of the microlocal spectrum condition.\\
Let us phrase this in more physical terms: A Hadamard two-point function(al) $\omega_{2}$ has only singularities with positive/negative frequencies w.r.t. to its first/second argument, while the causal propagator kernel $K_{G}$ has singularities with positive and negative frequencies equally contributing to both arguments. Nevertheless, the restriction of both distributions and their future normal derivatives to a Cauchy surface $\Sigma$ gives rise to \eqref{eq:hadamardbound}. But, the propagation of singularities is a problem of the dynamical law governing the identification of initial data and actual solutions (see \eqref{eq:symplecticspaces}), and the microlocal spectrum condition constrains the relevant state space subject to this evolution in a dynamical manner, which is only to a partial extent covered by \eqref{eq:hadamardbound}. In a sense, we may view this as an instance of Haag's theorem, which tells us that kinematical and dynamical aspects are tightly entangled in quantum field theory. This point is further elucidated by the fact that Hadamard states are suitable to define locally covariant, renormalized Wick products, time-ordered product and a stress-energy tensor, which are related to (perturbative) dynamical questions.\\
Therefore, let us have a short look at the dynamical law defined by $P$ in terms of initial data. Given two Cauchy surfaces $\Sigma_{1},\Sigma_{2}\subset M$, the causal propagator $G$ can be used to define a canonical transformation (see \eqref{eq:symplecticstructure})
\begin{align}
\label{eq:canonicalevolution}
\xymatrix@R=0.1cm{
\cD(\Sigma_{1},E_{\Sigma_{1}})^{\oplus2} \ar[r] & \cD(\Sigma_{2},E_{\Sigma_{2}})^{\oplus2} \\
(f^{\Sigma_{1}}_{0},f^{\Sigma_{1}}_{1}) \ar@{|->}[r] & \alpha^{G}_{\Sigma_{2},\Sigma_{1}}(f^{\Sigma_{1}}_{0},f^{\Sigma_{1}}_{1}):=(f^{\Sigma_{2}}_{0},f^{\Sigma_{2}}_{1})
}
\end{align}
with
\begin{align}
\label{eq:initialdataevolution}
f^{\Sigma_{2}}_{0} & =\iota^{*}_{\Sigma_{2}}((G'\circ(\iota_{\Sigma_{1}}^{*})')(f^{\Sigma_{1}}_{1})-(G'\circ(\nu_{\Sigma_{1}}^{*})')(f^{\Sigma_{1}}_{0})), \\
f^{\Sigma_{2}}_{1} & =\nu^{*}_{\Sigma_{2}}((G'\circ(\iota_{\Sigma_{1}}^{*})')(f^{\Sigma_{1}}_{1})-(G'\circ(\nu_{\Sigma_{1}}^{*})')(f^{\Sigma_{1}}_{0})).
\end{align}
This induces a *-isomorphism, also denoted by $\alpha^{G}_{\Sigma_{2},\Sigma_{1}}$, of the corresponding Weyl algebras by
\begin{align}
\xymatrix@R=0.1cm{
\cW_{\Sigma_{1}} \ar[r] & \cW_{\Sigma_{2}} \\
W(f^{\Sigma_{1}}_{0},f^{\Sigma_{1}}_{1}) \ar@{|->}[r] & \alpha^{G}_{\Sigma_{2},\Sigma_{1}}(W(f^{\Sigma_{1}}_{0},f^{\Sigma_{1}}_{1})):=W(f^{\Sigma_{2}}_{0},f^{\Sigma_{2}}_{1})
}
\end{align}
which can be pulled back to the their state spaces $(\alpha^{G}_{\Sigma_{2},\Sigma_{1}})^{*}:\mathcal{S}_{\Sigma_{1}}\rightarrow\mathcal{S}_{\Sigma_{2}}$. From and \eqref{eq:twopointsolutionfromdata} and \eqref{eq:twopointevaluation}, we infer that 
\begin{align}
\label{eq:twopointevolution}
((\alpha^{G}_{\Sigma_{2},\Sigma_{1}})^{*}\omega^{\Sigma_{2}})_{2,00}(f^{\Sigma_{1}}_{1},f'^{\Sigma_{1}}_{1}) & =\ \ \ \ \omega^{\Sigma_{2}}_{2,00}(\nu^{*}_{\Sigma_{2}}(G'\circ(\iota^{*}_{\Sigma_{1}})')(f^{\Sigma_{1}}_{1}),\nu^{*}_{\Sigma_{2}}(G'\circ(\iota^{*}_{\Sigma_{1}})')(f'^{\Sigma_{1}}_{1})) \\ \nonumber
 &\ \ \ \ -\omega^{\Sigma_{2}}_{2,01}(\nu^{*}_{\Sigma_{2}}(G'\circ(\iota^{*}_{\Sigma_{1}})')(f^{\Sigma_{1}}_{1}),\iota^{*}_{\Sigma_{2}}(G'\circ(\iota^{*}_{\Sigma_{1}})')(f'^{\Sigma_{1}}_{1})) \\ \nonumber
 &\ \ \ \ -\omega^{\Sigma_{2}}_{2,10}(\iota^{*}_{\Sigma_{2}}(G'\circ(\iota^{*}_{\Sigma_{1}})')(f^{\Sigma_{1}}_{1}),\nu^{*}_{\Sigma_{2}}(G'\circ(\iota^{*}_{\Sigma_{1}})')(f'^{\Sigma_{1}}_{1})) \\ \nonumber
 &\ \ \ \ +\omega^{\Sigma_{2}}_{2,00}(\iota^{*}_{\Sigma_{2}}(G'\circ(\iota^{*}_{\Sigma_{1}})')(f^{\Sigma_{1}}_{1}),\iota^{*}_{\Sigma_{2}}(G'\circ(\iota^{*}_{\Sigma_{1}})')(f'^{\Sigma_{1}}_{1}))\ \textup{etc}.
\end{align}
Thus, *-automorphisms of this form associated with two Cauchy surface are another way to state the correspondence between initial data and solutions. Now, we may say that Hadamard condition gives additional constraints on the initial data for two-point function(al)s, such that these data fit together via the dynamical law \eqref{eq:twopointevolution} to yield the positive/negative frequencies for the singularities. \\
In \cite{JunkerHadamardStatesAdiabatic, GerardConstructionOfHadamard}, we find explicit prescriptions in terms of pseudo-differential calculus how to construct and characterize Hadamard states on globally hyperbolic spacetimes, but these methods rely on the metric structure of the given spacetime, as we would expect, and therefore do not directly transfer to settings without such a structure, e.g. loop quantum gravity. More precisely, these construction use factorizations of the differential operator $P$ in terms of pseudo-differential operators to construct an explicit parametrization of Hadamard states. Interestingly, the construction in \cite{GerardConstructionOfHadamard} works with a characterization of Hadamard states in terms of optimal data as above, which is obtained from generic data by pullback with a pseudo-differential operator (see theorem 7.1 of \cite{GerardConstructionOfHadamard}). We observe, that the bound \eqref{eq:hadamardbound} is compatible with the conditions for the construction of a Hadamard state given in the said theorem. \\
In loop quantum gravity, such methods could only be applied in a certain semi-classical regime, where one reconstructs a spacetime metric $g$, or at least a spatial metric $q$ on $\Sigma$, from the geometric operators of the quantum theory. \\
To further elaborate on this point, let us consider a quantum algebra $\mathfrak{A}_{\Phi}$ of initial data on a Cauchy surface $\Sigma$ for a matter field on $M$, which is classically defined by a normally hyperbolic operator $P$ (or a quasi-linear version to include interactions), e.g. a Klein-Gordan field, a (gauge fixed) Maxwell-Yang-Mills field or a Higgs field. A state $\omega_{\Phi}:\mathfrak{A}_{\Phi}\rightarrow\C$ of the quantum field may or may not depend on the spacetime metric $g$ or its restriction $q$ to $\Sigma$, e.g. a Hadamard state $\omega_{H}$ in the first case, or a background independent state based on a cylindrical measure for functions of point-holonomies ``$\exp(i\Phi)$'' in the second case (cf. \cite{ThiemannKinematicalHilbertSpaces, ThiemannQuantumSpinDynamics5}). The second possibility is what we expect to happen in loop quantum gravity, or any theory, where a (classical) spacetime metric is not directly available in the quantum theory. In such cases, an application of the microlocal spectrum condition to quantum states of matter will require the restriction of the theory to some sort of semi-classical sector, which provides us with an effective spacetime metric $g_{\textup{eff}}$. Assuming this, we can ask whether the quantum state $\omega_{\Phi}$ is, at least approximately, Hadamard w.r.t. $g_{\textup{eff}}$. Clearly, we can only expect the microlocal spectrum condition to be satisfied in an approximate sense, if we use irregular states for the quantum matter field as proposed in \cite{ThiemannKinematicalHilbertSpaces}, because the microlocal spectrum condition requires the existence of the two-point function(al) of the state $\omega_{\Phi}$. A detailed discussion of the latter issue for the Maxwell field, quantized by loop quantum gravity methods, will be presented elsewhere. Nevertheless, as an example, we take a look at the so-called polymer representation of a linear scalar field, $\phi$, which is obtained from the following irregular state on the Weyl algebra,
\begin{align}
\label{eq:pointholalg}
W(\lambda\delta_{x},f) & = e^{\frac{i}{2}\lambda\delta_{x}(f)}U(x,\lambda)V(f), \\ \nonumber
W(\lambda\delta_{x},f)W(\lambda’\delta_{x’},f’) & = e^{-\frac{i}{2}(\delta_{x}(f’)-\delta_{x’}(f))}W(\lambda\delta_{x}+\lambda’\delta_{x’},f+f’), \\ \nonumber
W(\lambda\delta_{x},f)^{*} & = W(-\lambda\delta_{x},-f),
\end{align}
generated by point-holonomies, $U(x,\lambda) = e^{i\lambda\delta_{x}(\phi)} = e^{i\lambda\phi(x)},\ x\in\Sigma,\lambda\in\R$, and exponentials of smeared momenta, $V(f) = e^{i\pi(f)},\ f\in C^{\infty}_{c}(\Sigma),$ (cf. \cite{ThiemannKinematicalHilbertSpaces, AshtekarPolymerAndFock}, see also \cite{ThiemannModernCanonicalQuantum}, Chapter 12):
\begin{align}
\label{eq:polymerrep}
\omega_{0}\left(W\left(\sum_{i=1}^{m}\lambda_{i}\delta_{x_{i}},f\right)\right) & = \prod_{i=1}^{m}\delta_{\lambda_{i},0},\ \ \ m\in\N, \lambda_{1},...,\lambda_{m}\in\R, f\in C^{\infty}_{c}(\Sigma),
\end{align}
where $x_{1},...,x_{m}$ are mutually distinct. This state resembles the Ashtekar-Isham-Lewandowski state used in loop quantum gravity. As the point fields, $\phi(x),\ x\in\Sigma$, are not well-defined in the polymer representation because of the irregularity of $\omega_{0}$, we need to introduce regularised substitutes to apply our condition to the initial data of the two-point function of $\omega_{0}$, at least in an approximate sense. To this end, we define
\begin{align}
\label{eq:lqcpointfield}
\phi_{\lambda}(x) & := \tfrac{1}{2i\lambda}\left(U(x,\lambda)-U(x,\lambda)^{*}\right),\ \ \ \lambda>0,
\end{align}
in analogy with loop quantum cosmology (cf. \cite{AshtekarLoopQuantumCosmology}). Using the \eqref{eq:pointholalg}, we find:
\begin{align}
\label{eq:twopointpolymer}
\omega_{0}\left(\phi_{\lambda}(x)\pi(f)\right) & = \frac{i}{2}f(x)\delta_{\lambda,0} \underset{\lambda>0}{=} 0, & \omega_{0}\left(\pi(f)\phi_{\lambda}(x)\right) & = -\frac{i}{2}f(x)\delta_{\lambda,0} \underset{\lambda>0}{=} 0, \\ \nonumber
\omega_{0}\left(\phi_{\lambda}(x)\phi_{\lambda’}(x’)\right) & = \frac{1}{2\lambda^{2}}(\delta_{\lambda,\lambda’}+\delta_{\lambda,-\lambda’})\delta_{x,x’}, & \omega_{0}(\pi(f)\pi(f’)) & = 0.
\end{align}
Hence, the wave front sets of the (regularised) initial data of the two-point function of $\omega_{0}$ are empty, and our condition does not apply. But, since irregular states like $\omega_{0}$ are typically used in approaches like loop quantum gravity, in which spacetime is expected to be of a granular, discrete nature, the immediate application of our condition might be bound to fail anyhow, because the Hadamard condition is a statement about the ultra-short distance behaviour of the two-point function, becoming meaningless in such contexts.\\
Indeed, to make sense of a condition on the ultra-short distance behaviour of a quantum field’s two-point function, it might be necessary to introduce a coarse graining procedure that allows us to recover an effective continuum theory describing the quantum field. Some intriguing work in this direction can be found in \cite{VaradarajanFockRepresentationsFrom, VaradarajanPhotonsFromQuantized, AshtekarPolymerAndFock} and be fitted into our example:\\[0.1cm]
In the case $\Sigma = \R^{3}$, we define a $\varepsilon$-scale of Fock states,
\begin{align}
\label{eq:fockscale}
\omega_{F,\varepsilon}(W(f,f’)) & = e^{-\frac{1}{4}\left(\varepsilon^{-1}||f||_{L^{2}}^{2}+\varepsilon||f’||_{L^{2}}^{2}\right)},\ \ \ f,f’\in C^{\infty}_{c}(\R^{3}),
\end{align}
together with an $r$-scale of coarse graining maps (algebra isomorphisms)
\begin{align}
\label{eq:coarsegrain}
\alpha_{r}:\ W(\lambda\delta_{x},K_{r}\ast f) & \longmapsto W(\lambda K_{r}\circ\tau_{-x},f)
\end{align}
along flat background metrics. Here, $K_{r}(x) = \frac{1}{\sqrt{2\pi r}^{3}} e^{-\frac{|x|^{2}}{2r}}$ is the heat kernel on $\R^{3}$ at time $r$, and $\tau_{-x}(x’) = x’-x$ is the translation map. The regularised initial data of $\omega_{F,\varepsilon}$ take the form:
\begin{align}
\label{eq:twopointfock}
\omega_{F,\varepsilon}(\alpha_{r}(\phi_{\lambda}(x)\pi(f))) & = \frac{i}{2}(K_{r}\ast f)(x)e^{-\frac{\lambda^{2}}{4\varepsilon}\sqrt{4\pi r}^{-3}}, \\ \nonumber
\omega_{F,\varepsilon}(\alpha_{r}(\pi(f)\phi_{\lambda}(x))) & = -\frac{i}{2}(K_{r}\ast f)(x)e^{-\frac{\lambda^{2}}{4\varepsilon}\sqrt{4\pi r}^{-3}} \\ \nonumber
\omega_{F,\varepsilon}(\alpha_{r}(\phi_{\lambda}(x)\phi_{\lambda’}(x’))) & = \tfrac{1}{\lambda\lambda’}\sinh\left(\tfrac{\lambda\lambda’}{2\varepsilon}(K_{r}\circ\tau_{-x},K_{r}\circ\tau_{-x’})_{L^{2}}\right)e^{-\frac{\lambda^{2}+(\lambda’)^{2}}{4\varepsilon}\sqrt{4\pi r}^{-3}}, \\ \nonumber \omega_{F,\varepsilon}(\alpha_{r}(\pi(f)\pi(f’))) & = \frac{\varepsilon}{2}(f,f’)_{L^{2}}.
\end{align}
In limit $\varepsilon\rightarrow0+$, succeeded by $r\rightarrow0+$, we recover \eqref{eq:twopointpolymer}, which is consistent with $\omega_{F,\varepsilon}\circ\alpha_{r}\rightarrow\omega_{0}$, while in the limit $\lambda,\lambda’\rightarrow0+$, followed by $r\rightarrow0+$, we have:
\begin{align}
\label{eq:twopointfocklimit}
\omega_{F,\varepsilon}(\phi(x)\pi(x’)) & = \frac{i}{2}\delta_{x}(x’), & \omega_{F,\varepsilon}(\pi(x)\phi(x’)) & = -\frac{i}{2}\delta_{x}(x’), \\ \nonumber
\omega_{F,\varepsilon}(\phi(x)\phi(x’)) & = \frac{1}{2\varepsilon}\delta_{x}(x’), & \omega_{F,\varepsilon}(\pi(x)\pi(x’)) & = \frac{\varepsilon}{2}\delta_{x}(x’).
\end{align}
The latter are consistent with our condition on initial data of a Hadamard state’s two-point function, but could only be obtained because the $\varepsilon$-scaled Fock states are sufficiently regular to allow for the existence of the point fields $\phi(x),\ x\in\Sigma,$ in a distributional sense.\\
A proposal by the authors for the construction of a semi-classical sector within (canonical) loop quantum gravity, which roughly follows ideas presented in \cite{SahlmannTowardsTheQFT1, SahlmannTowardsTheQFT2, GieselBornOppenheimerDecomposition}, will be put forward soon \cite{StottmeisterCoherentStatesQuantumI, StottmeisterCoherentStatesQuantumII, StottmeisterCoherentStatesQuantumIII}. Methods that achieve this in symmetry reduced models, i.e. loop quantum cosmology, and make contact with the theory of adiabatic vacua, have already been established and been applied to cosmological perturbation theory \cite{AgulloThePreInflationary, AgulloExtensionOfThe}.\\
Altenatively, one could try to adapt the factorization techniques to deparametrizing models and their quantum Hamiltonians (see \cite{GieselScalarMaterialReference}) to find analogs of the positive/negative frequency condition.\\
Finally, let us point out that the microlocal spectrum condition is tailored to (linear) quantum fields defined by normally hyperbolic operators, because the characteristic set of such operators is the co-light cone bundle of $(M,g)$, a fact that could remain true only to zeroth oder in $l_{\textup{Planck}}$ in quantum gravity.\\
Nevertheless, let us also note that a Hadamard state allows for the perturbative construction of interactions, because Wick products and time-ordered products exist \cite{BrunettiMicrolocalAnalysisAnd, HollandsLocalWickPolynomials, HollandsExistenceOfLocal}, due to the fact that the products of distributions $\omega_{2}(x,y)^{n}, n\in\N,$ are well-defined by H\"ormander's theorem \ref{thm:distproduct}. Although, distributions of this form, e.g. $:\Phi^{n}:$, will no longer satisfy the microlocal spectrum condition, there are generalized bounds on their wave front sets \cite{BrunettiMicrolocalAnalysisAnd}. But, a restriction of these product distributions to a Cauchy surface does in general not exist (Schwinger terms).
\section{Acknowledgements}
\label{sec:ack}
AS gratefully acknowledges financial support by the Ev. Studienwerk e.V.. This work was supported in parts by funds from the Friedrich-Alexander-University, in the context of its Emerging Field Initiative, to the Emerging Field Project ``Quantum Geometry’’.

\section{Appendix}
\label{sec:app}
This appendix is intended to provide some mathematical background material and key results from microlocal analysis (cf. \cite{HoermanderTheAnalysisOf1}, see also \cite{JunkerHadamardStatesAdiabatic, SahlmannMicrolocalSpectrumCondition, DabrowskiFunctionalPropertiesOf}). The definitions for distributions on manifolds follow those in \cite{BaerWaveEquationsOn} (see also \cite{FriedlanderTheWaveEquation, RudinFunctionalAnalysis, WernerFunktionalanalysis}).

\subsection{Distributions on manifolds}
\label{sec:distmanifold}
Let $M$ be a finite dimensional, Hausdorff, second-countable, $\sigma$-compact\footnote{This means, there is a countable exhaustion $\{K_{n}\}_{n=1}^{\infty}$ of $M$ by compact sets $K_{n}\sqsubset M$.}, smooth manifold. Given a vector bundle $E\stackrel{\pi}{\rightarrow}M$,we denote by $\cD(M,E):=\Gamma^{\infty}_{0}(M,E)$ the space of smooth, compactly supported sections. This space can be made into a nuclear, strict LF-space (see the explanation following \eqref{eq:testsectionclosedsubspaces}) by the following semi-norms:\\[0.1cm]
Fix an arbitrary Riemmanian metric $g$ on $M$ and an arbitrary fibre metric $g_{E}$ (hermitian in the complex case) on $E$. Additionally, choose arbitrary connections in $T^{*}M$ and $E$, such that we have induced connections\footnote{All connections are denoted by the same symbol $\nabla$.} $\nabla:\Gamma^{\infty}(M,T^{*}M^{\otimes k}\otimes E)\rightarrow\Gamma^{\infty}(T^{*}M^{\otimes k+1}\otimes E),\ \forall k\N_{0}$. The metrics $g,g_{E}$ induce norms $||\ .\ ||_{g,g_{E}}:T^{*}M^{\otimes k}\otimes E\rightarrow\R_{\geq0},\ \forall k\in\N_{0}$, and we define for $f\in\cD(M,E)$:
\begin{align}
\label{eq:NLFnorms}
||f||_{C^{n}(K,E)} & := \max_{j=1,...,n}\sup_{x\in K}||\nabla^{j}f(x)||_{g,g_{E}},\ n\in\N_{0},K\sqsubset M\ \textup{compact}.
\end{align}
Clearly, different choices of metrics and connections lead to equivalent semi-norms, because $K$ is compact. Now, we introduce the spaces:
\begin{align}
\label{eq:testsectionclosedsubspaces}
\cD_{K}(M,E) & :=\{f\in\cD(M,E)\ | \supp(f)\in K\},\ K\sqsubset M\ \textup{compact},
\end{align}
which we turn into Frech\'et spaces with the families of semi-norms $\{||\ .\ ||_{C^{n}(K,E)}\}_{n\in\N}$. The nuclear, strict LF-topology on $\cD(M,E)$ is defined as the topology generated by all semi-norms $p:\cD(M,E)\rightarrow\R_{\geq0}$, s.t. all the restrictions $p_{|\cD_{K}(M,E)},\ K\sqsubset M\ \textup{compact},$ are continuous (in $\cD_{K}(M,E)$). This topology has the important property that it turns $\cD(M,E)$ into a barreled space (in which the Banach-Steinhaus theorem or principle of uniform boundedness holds, cf. \cite{RobertsonTopologicalVectorSpaces}), and entails the following notions of convergence in $\cD(M,E)$:
\begin{proposition}
\label{prop:testsectionconvergence}
A sequence $\{f_{j}\}_{j=1}^{\infty}\subset\cD(M,E)$ converges to $f\in\cD(M,E)$, if and only if
\begin{itemize}
	\item[1.] $\exists K\sqsubset M\ \textup{compact}:\ \forall j:\ \supp(f_{j}),\supp(f)\subset K$,
	\item[2.] $\forall n\in\N_{0}:\ \lim_{j\rightarrow\infty}||f_{j}-f||_{C^{n}(K,E)}=0$.
\end{itemize}
\end{proposition}
$\cD(M,E)$ with its nuclear, strict LF-topology is called the \textit{space of test section in} $E$ \textit{on} $M$. Distributions in $E^{*}$\footnote{$E^{*}$ is the fibrewise dual of $E$.} on $M$ with values in a real or complex, finite dimensional vector space $V$ can be defined as sequentially continuous maps $\cD(M,E)\rightarrow V$, where we fix some arbitrary norm $||\ .\ ||_{V}$ on $V$.
\begin{definition}
\label{def:distmanifold}
We denote the space of sequentially continuous maps $\cD(M,E)\rightarrow V$ endowed with the weak$^{*}$-topology is by $\cD'(M,E^{*},V)$, and call it the \textit{space of distributions in} $E^{*}$ \textit{on} $M$ \textit{with values in} $V$. If $V=\R,\C$, we abbreviate the notation by $\cD'(M,E^{*})$.
\end{definition}
Equivalently, we can characterize distributions in the following way.
\begin{proposition}
\label{prop:distorder}
For a map $u:\cD(M,E)\rightarrow V$ the following conditions are equivalent:
\begin{itemize}
	\item[1.] $u\in\cD'(M,E^{*},V)$,
	\item[2.] $\forall K\sqsubset M\ \textup{compact}\ \exists k\in\N_{0}, \infty>C>0:\ \forall f\in\cD(M,E):\ ||u(f)||_{V}\leq||f||_{C^{k}(K,E)}$.
\end{itemize}
\end{proposition}
If $M$ is orientable, we may choose a (smooth) volume form $dV$ on it\footnote{More generally, we can use a nowhere vanishing density on $M$.}. This gives rise to a continuous embedding:
\begin{align}
\label{eq:testsectionembedding}
\xymatrix@R=0.1cm{
\cD(M,E^{*})\ \ar@{^{(}->}[r] & \cD'(M,E^{*}) \\
f \ar@{|->}[r] & \left(f'\mapsto\int_{M}(f,f')dV=:u_{f,dV}(f')\right)
}
\end{align}
where $f'\in\cD(M,E)$. Since two volume forms $dV,dV'$ differ by a nowhere vanishing function $f_{dV,dV'}\in C^{\infty}(M)$, any two embeddings of this kind are equivalent. This motivates the definition of derivatives of distributions.
\begin{definition}
\label{def:distderivative}
A linear differential operator $P:\Gamma^{\infty}(M,E)\rightarrow\Gamma^{\infty}(M,E)$ uniquely extends to a continuous, linear operator $P':\cD'(M,E^{*})\rightarrow\cD'(M,E^{*})$ by
\begin{align}
\label{eq:distderivative}
\forall u\in\cD'(M,E^{*}):\ (P'u)(f) & := u(Pf),\ f\in\cD(M,E).
\end{align}
\end{definition}
Equation \eqref{eq:distderivative} is compatible with the definition of the \textit{formal adjoint} $P^{*}:\Gamma^{\infty}(M,E^{*})\rightarrow\Gamma^{\infty}(M,E^{*})$ \textit{of} $P$ \textit{relative to} $dV$, because of the identity:
\begin{align}
\label{eq:formaladjoint}
\int_{M}(P^{*}f,f')dV & = \int_{M}(f,Pf')dV,\ f\in\cD(M,E^{*}),f'\in\cD(M,E).
\end{align}
Next, we define the \textit{support of a distribution}, as the generalization of the support of a function resp. section..
\begin{definition}
\label{def:distsupport}
The support $\supp(u)$ of a distribution $u\in\cD'(M,E^{*},V)$ is the complement of the set
\begin{align}
\label{eq:distsupport}
\{x\in M\ |\ \exists U\subset M\ \textup{open},x\in U:\ u(f)=0\ \forall f\in\cD(M,E), \supp(f)\subset U\}.
\end{align}
Clearly, $\supp(u)$ is closed in $M$.
\end{definition}
The distributions with compact support $\cD'_{0}(M,E^{*})$ can be considered as the (distributional) dual of the smooth section in $E$, $\Gamma^{\infty}(M,E)$, because of the identity:
\begin{align}
\label{eq:distcompactsupportidentity}
u(f) & = u(\varphi f),
\end{align}
where $\varphi\in\cD(M)$ is a test function with $\varphi\equiv1$ on a neighborhood of $\supp(u)$. We can turn the smooth sections $\Gamma^{\infty}(M,E)$ into a nuclear Frech\'et space $\cE(M,E)$ by the semi-norms \eqref{eq:NLFnorms}, s.t. its weak$^{*}$-topological, $V$-valued dual is the space of distributions with compact support $\cE'(M,E^{*},V)=\cD'_{0}(M,E^{*},V)$. The spaces $\cD_{K}(M,E)$ are closed subspaces of $\cE(M,E)$. The notion of convergence in $\cE(M,E)$ is given by:
\begin{proposition}
\label{prop:smoothsectionconvergence}
A sequence $\{f_{j}\}_{j=1}^{\infty}\subset\cE(M,E)$ converges to $f\in\cE(M,E)$, if and only if 
\begin{align}
\label{eq:smoothsectionconvergence}
\forall K\sqsubset M\ \textup{compact},\ n\in\N_{0}:\ \lim_{j\rightarrow\infty}||f_{j}-f||_{C^{n}(K,E)}=0.
\end{align}
\end{proposition}
There is a characterization of the elements in $\cE'(M,E^{*},V)$ similar to proposition \ref{prop:distorder}, as well.
\begin{proposition}
\label{prop:compactdistorder}
For a map $u:\cE(M,E)\rightarrow V$ the following conditions are equivalent:
\begin{itemize}
	\item[1.] $u\in\cE'(M,E^{*},V)$,
	\item[2.] $\exists K\sqsubset M\ \textup{compact},\ k\in\N_{0}, \infty>C>0:\ \forall f\in\cD(M,E):\ ||u(f)||_{V}\leq||f||_{C^{k}(K,E)}$.
\end{itemize}
\end{proposition}

\subsection{The wavefront set - Tools from microlocal analysis}
\label{sec:microlocalanalysis}
A main advantage in the theory of distribution on $\R^{n}$ is the applicability of the Fourier transform to investigate smoothness properties. This can be, at least partly, cast into a local notion generalizable to ($C^{\infty}$-)manifolds, namely the so-called \textit{wave front set}
\begin{equation}
\label{eq:wavefrontset}
\WF(u)\subset\textup{T*M}\setminus\{0\},\ u\in\cD'\left(M\right)
\end{equation}
which will be as indicated a subset of the cotangent bundle of $M$. This set captures information on the (co-)directions along which the singularities of $u$ ``\textit{propagate}'', and e.g. allows for a refined analysis of the operations possible with distributions.\\[0.25cm]
To define the wave front set explicitly we need the following ``localization'' of the decay properties of distributions on $\R^{n}$:
\begin{definition}
\label{def:localizingdist}
For $u\in\cD'\left(\R^{n}\right)$ we call $(x,k)\in\R^{n}\times\left(\R^{n}\setminus\{0\}\right)$ a \textup{regular direction} of $u$ at $x$, if there exists $\phi\in\cD\left(\R^{n}\right)$ with $\phi(x)\neq0$ and a \textup{conic} open neighborhood $\Gamma\subset\R^{n}\setminus\{0\}$ of $k$, s.t.
\begin{equation}
\label{eq:localizingdist}
\forall N\in\N:\ \sup_{k'\in\Gamma}(1+|k'|)^{N}|\widehat{\phi u}(k')|\leq C_{N}<\infty
\end{equation}
Recall that a set $\Gamma$ is called conic if $k\in\Gamma\Leftrightarrow r k\in\Gamma,\ r\in\R_{>0}$.\\[0.1cm]
Let $\Sigma_{x}(u)$ denote the complement of the regular directions at $x$.
\end{definition}
We observe that this definition is local in the sense that $\supp\phi$ can arbitrarily concentrated around $x$, i.e. $\forall\phi\in\cD\left(\R^{n}\right):\phi(x)\neq0:\Sigma_{x}(\phi u)=\Sigma_{x}(u)$.
\begin{definition}
\label{def:wavefrontset}
The \textup{wave front set} of $u\in\cD'\left(\R^{n}\right)$ is given by the set
\begin{equation}
\label{eq:wavefrontsetdef}
\WF(u)=\left\{(x,k)\in\R^{n}\times\left(\R^{n}\setminus\{0\}\right)\ |\ k\in\Sigma_{x}\right\}
\end{equation}
Clearly, $\WF(u)$ is conic in the sense that it is invariant under multiplication of the second component by positive scalars, i.e. $(x,k)\in\WF(u)\Leftrightarrow(x,r k)\in\WF(u), r>0$.
\end{definition}
An immediate consequence of the definition is that the wave front set naturally generalizes the notion of singular support of a distribution.
\begin{corollary}
\label{cor:projectionsingsupp}
The projection of $\WF(u)$ onto the first component is $\ssupp u$.
\end{corollary}
Another observation following from the interplay of the Fourier transform and the complex conjugation is:
\begin{corollary}
\label{cor:conjugatewavefrontset}
For $u\in\cD'\left(\R^{n}\right)$ one has
\begin{equation}
\label{eq:conjugatewavefrontset}
\WF(\bar{u})=\left\{(x,k)\in\R\times\left(\R\setminus\{0\}\right)\ |\ (x,-k)\in\WF(u)\right\}=-\WF(u),
\end{equation}
where $\bar{u}$ denotes the complex conjugate.
\end{corollary}
Moreover, in analogy with the support of a distribution, we have the following local behavior of the wave front set.
\begin{corollary}
\label{cor:differentialwavefrontset}
For any linear ($C^{\infty}$-)differential operator $P$ the wave front set has the property
\begin{equation}
\label{eqn:differentialwavefrontset}
\WF(Pu)\subset\WF(u)
\end{equation}
\end{corollary}
To realize the wave front set as part of the cotangent bundle of a manifold one needs its transformation behavior under ($C^{\infty}$-)maps $\Phi:U\subset\R^{n}\longrightarrow V\subset\R^{m}$ between open sets.
\begin{definition}
\label{def:conormalset}
The \textup{co-normal} of $\Phi$ is the set
\begin{equation}
\label{eq:conormalset}
N_{\Phi}=\left\{(\Phi(x),\eta)\in V\times\R^{m}\ |\ (d\Phi_{x})^{*}\eta = 0\right\}.
\end{equation} 
Obviously we have $N_{\Phi}=\{0\}$ if $\Phi$ is a submersion, i.e. $d\Phi$ is everywhere onto.
\end{definition}
As the main obstacle in defining the composition of distributions with ($C^{\infty}$-)maps is due to the presence of singularities one is led to consider spaces of distributions with certain restrictions on their wave front set. This paves the way to extending operations (e.g. multiplication) from ($C^{\infty}$-)functions to distributions.
\begin{definition}[H\"ormander's pseudo-topology]
\label{def:restrictedwavefrontsetdist}
For an open subset $U\subset\R^{n}$ and a closed cone $\Gamma\subset U\times\left(\R^{n}\setminus\{0\}\right)$ consider the set
\begin{equation}
\label{eq:restrictedwavefrontsetdist}
\cD'_{\Gamma}(U)=\left\{u\in\cD'(U)\ |\ \WF(u)\subset\Gamma\right\}.
\end{equation}
A sequence $\left\{u_{i}\right\}\subset\cD'_{\Gamma}(U)$ is said to converge to $u\in\cD'_{\Gamma}(U)\ (u_{i}\rightarrow u)$ within $\cD'_{\Gamma}(U)$ if\footnote{This entails that $\cD(U)\subset\cD'_{\Gamma}(U)$ is a dense subset.}
\begin{itemize}
	\item[\textup{(i)}]  $u_{i}\rightarrow u$ in $\cD'(U)$
	\item[\textup{(ii)}] $\forall N\in\N:\forall\phi\in\cD(U):\forall V\subset\R^{n}\ \textup{closed cone}:\Gamma\cap\left				    (\supp\phi\times V\right)=\emptyset$
\begin{equation}
\label{eq:restrictedwavefrontsetconvergence}
\sup_{V}|k|^{N}|\widehat{\phi u_{i}}(k)-\widehat{\phi u}(k)|\rightarrow0,\ i\rightarrow\infty.
\end{equation}
\end{itemize}
\end{definition}
There are several topologies on $\cD'_{\Gamma}(U)$ compatible with this notion of convergence (cf. \cite{DabrowskiFunctionalPropertiesOf}). Now we are in the position to state the main theorem:
\begin{theorem}[Theorem 8.2.4. \cite{HoermanderTheAnalysisOf1}]
\label{thm:distpullback}
Let  $\Phi:U\subset\R^{n}\longrightarrow V\subset\R^{m}$ be as above. There is one and only one way to define the \textup{pullback} $\Phi^{*}u$ for $u\in\cD'\left(\R^{n}\right)$ with
\begin{equation}
\label{eq:conormalintersection}
N_{\Phi}\cap\WF(u)=\emptyset
\end{equation}
such that $\Phi^{*}u=u\circ\Phi$ for $u\in C^{\infty}$, and for any closed conic subset $\Gamma\subset V\times\left(\R^{m}\setminus\{0\}\right)$ with $\Gamma\cap N_{\Phi}=\emptyset$
\begin{equation}
\label{eq:distpullback}
\Phi^{*}:\cD'_{\Gamma}\left(V\right)\longrightarrow\cD'_{\Phi^{*}\Gamma}\left(U\right)
\end{equation}
$\Phi^{*}\Gamma=\left\{(x,(d\Phi_{x})^{*}\eta)\ |\ (\Phi(x),\eta)\in\Gamma\right\}$ is continuous.\\[0.1cm]
Moreover the wave front set satisfies
\begin{equation}
\label{eq:wavefrontsetpullback}
\WF(\Phi^{*}u)\subset\Phi^{*}WF(u)
\end{equation}
\end{theorem}
Interestingly this makes precise the intuition that the singularities of a distribution (as a geometrical object on a manifold) should ``propagate'' along tangential direction and not along the co-normal.\\[0.25cm]
Consider now a ($C^{\infty}$-)manifold $M$ and a distribution $u\in\cD'(M)$. Utilizing theorem \ref{thm:distpullback} we define the wave front set $\WF(u)\subset\textup{T*M}\setminus\{0\}$ by
\begin{equation}
\label{eq:manifoldwavefrontset}
(x,k)\in\WF(u)\Leftrightarrow(\kappa(x),(d\kappa^{-1}_{x})^{*}k)\in\WF((\kappa^{-1})^{*}u)
\end{equation}
for any chart $\kappa:U\subset M\longrightarrow V\subset\R^{n}$. In case of a ($C^{\infty}$-)vector bundle $E$ over $M$ and $u\in\cD'(M,E)$ one defines $\WF(u):=\bigcup_{i=1,...,e}\WF(u_{i})$ w.r.t. a local trivializations s.t. ($u=(u_{1},...,u_{e})$). This is independent of the trivialization since the passage between two trivialization is given by the multiplication of $(u_{1},...,u_{e})$ by an invertible ($C^{\infty}$-)matrix.\\[0.25cm]
Another important implication of theorem \ref{thm:distpullback} is the possibility to define restrictions of distributions to submanifolds in certain cases:
\begin{corollary}
\label{cor:submanifoldwavefrontset}
Let $\iota:S\longrightarrow M$ be a(n) (embedded) submanifold. For every $u\in\cD'(M)$ with $\WF(u)\cap N_{\iota}=\emptyset$ the restriction
\begin{equation}
\label{eq:submanifolddist}
u_{|S}=\iota^{*}u
\end{equation}
is a well defined distribution in $S$ $(u_{|S}\in\cD'(S))$.
\end{corollary}
Next we take a closer look at the wave front set of the (exterior) tensor product of distributions, which will be important due to the fact, that the product of ($C^{\infty}$-)functions can be given as
\begin{equation}
\label{eq:productaspullback}
fg(x)=\Delta^{*}(f\otimes g)(x),
\end{equation}
where $\Delta:M\longrightarrow M\times M$ is the diagonal map.
\begin{proposition}
\label{prop:tensorproductwavefrontset}
For $u\in\cD'(M)$, $v\in\cD'(M')$ the wave front set of the (exterior) tensor product $u\otimes v\in\cD'(M\times M')$ obeys the restriction
\begin{equation}
\label{eq:tensorproductwavefrontset}
\WF(u\otimes v)\subset\left(\WF(u)\times\WF(v)\right)\cup\left((\supp u\times\{0\})\times\WF(v)\right)\cup\left(\WF(u)\times(\supp v\times\{0\})\right).
\end{equation}
\begin{proof}
The Fourier transform of $(\phi u)\otimes(\psi v)$ is given by $\widehat{\phi u}\widehat{\psi v}$ ($\phi(x)\neq0,\ \psi(y)\neq0$). According to relation \eqref{eq:localizingdist} we have for the regular directions at $(x,y)$ (w.r.t. to a local coordinate system):
\begin{equation}
\label{eq:localizingproductdist}
\forall N\in\N:\ \sup_{(k,k')\in\Gamma\subset(T^{*}_{x}M\times T^{*}_{y}M')\setminus\{0\}}(1+\underbrace{|(k,k')|}_{=|k|+|k'|})^{N}|\widehat{\phi u}(k)\widehat{\psi v}(k')|\leq C_{N}<\infty.
\end{equation}
So we infer that
\begin{equation}
\label{eq:decayproductdist}
\Sigma_{x}(u\otimes v)\subset\left(\Sigma_{x}(u)\times\Sigma_{y}(v)\right)\cup\left(\{0\}\times\Sigma_{y}(v)\right)\cup\left(\Sigma_{x}(u)\times\{0\}\right),
\end{equation}
which implies the result. \qed
\end{proof}
\end{proposition}
Obviously, the co-normal of the diagonal map $\Delta$ is given by
\begin{equation}
\label{eq:diagonalconormalset}
N_{\Delta}=\left\{(\Delta(x),(k,l))\in\textup{T*M}^{\times2}\ |\ k=-l\right\},
\end{equation}
leading together with proposition \ref{prop:tensorproductwavefrontset} to the extension theorem for multiplication:
\begin{theorem}[Theorem 8.2.10. \cite{HoermanderTheAnalysisOf1}]
\label{thm:distproduct}
For $u,v\in\cD'(M)$ the product $uv$ is well-defined if $\WF(u\otimes v)\cap N_{\Delta}=\emptyset$, i.e. there is no $(x,k)\in\textup{T*M}$ s.t. $(x,k)\in\WF(u)$ and $(x,-k)\in\WF(v)$, and given by
\begin{equation}
\label{eq:distproduct}
uv = \Delta^{*}(u\otimes v).
\end{equation}
Furthermore the wave front $\WF(uv)$ satisfies
\begin{equation}
\label{eq:distproductwavefrontset}
\WF(uv)\subset\WF(u)\oplus\WF(v)\cup\WF(u)_{|\supp v}\cup\WF(v)_{|\supp u}.
\end{equation}
\end{theorem}
Another class of important theorems concerns the composition of distributions as linear maps. The first is essentially a refined version of theorem 8.2.12. in \cite{HoermanderTheAnalysisOf1}.
\begin{theorem}
\label{thm:distkernelcomposition}
Let $U\subset\R^{n}$ and $V\subset\R^{m}$ be open sets and $K\in\cD'(U\times V)$. Denote by $\mathcal{K}$ the corresponding map $\cD(V)\longrightarrow\cD'(U)$. Then the wave front set $\WF(\mathcal{K}\psi)$ for $\psi\in\cD(V)$ satisfies
\begin{align}
\label{eq:wavefrontsetkernelmap}
\ &\ \{(x,k)\in U\times\left(\R^{n}\setminus\{0\}\right)\ |\ (x,y;k,0)\in\WF(K),\ y\in\supp^{\circ}\psi\} \\[0.2cm]
\subset\ &\ \WF(\mathcal{K}\psi) \\[0.2cm]
\subset\ &\ \{(x,k)\in U\times\left(\R^{n}\setminus\{0\}\right)\ |\ (x,y;k,0)\in\WF(K),\ y\in\supp \psi\},
\end{align}
where $^{\circ}$ denotes the interior of a set. Defining
\begin{equation}
\label{eq:projectedwavefrontset}
\WF(K)_{U|V}=\left\{(x,k)\in U\times\left(\R^{n}\setminus\{0\}\right)\ |\ (x,y;k,0)\in\WF(K),\ y\in V\right\}
\end{equation}
one has
\begin{equation}
\label{eq:projectedwavefrontsetkernelmap}
\WF(K)_{U|\supp^{\circ}\psi}\subset\WF(\mathcal{K}\psi)\subset\WF(K)_{U|\supp \psi}.
\end{equation}
\begin{proof}
For $x_{0}\in U$ take $\phi\in\cD(U)$ with $\phi(x_{0})=1$ and define
\begin{equation}
\label{eq:localizedkernel}
K_{1}=(\phi\otimes\psi)K\in\cE'(U\times V),\ \psi\in\cD(V).
\end{equation}
To analyze the wave front set we look at $\widehat{\phi(\mathcal{K}\psi)}$:
\begin{equation}
\label{eq:fourierlocalizedkernel}
\widehat{\phi(\mathcal{K}\psi)}(k) = \phi(\mathcal{K}\psi)\left(e^{-i k\cdot(\ .\ )}\right) = K\left(\phi e^{-i k\cdot(\ .\ ),\psi}\right) = \widehat{K_{1}}(k,0).
\end{equation}
If $u\in\cE'(\R^{p})$ is a compactly supported distribution we denote by $\Sigma(u)$ the complement of the regular directions of its Fourier transform, s.t. $\pi_{2}(\WF(u))=\Sigma(u)$\footnote{$\pi_{i},\ i=1,2$ denotes the projection on the respective component.}. Moreover one can show\footnote{see \cite{HoermanderTheAnalysisOf1} p. 253 et seq.} for $\chi\in\cD(\R^{p})$
\begin{equation}
\Sigma(\chi u)\subset\Sigma(u)\ \textup{and}\ \Sigma(\chi u)\rightarrow\Sigma_{x}(u)\ \textup{for}\ \supp\chi\rightarrow\{x\},\ \chi(x)\neq0.
\end{equation}
This directly leads to $\bigcup_{(x,y)\in\supp^{\circ}\phi\otimes\psi}\Sigma_{(x,y)}(K)\subset\Sigma(K_{1})$ and $\Sigma_{(x,y)}(K_{1})\subset\Sigma_{(x,y)}(K)$. So we find:
\begin{align}
\ &\ \pi_{2}\left(\WF(K)_{|\supp^{\circ}\phi\otimes\psi}\right)\subset\Sigma(K_{1})=\pi_{2}\left(\WF(K_{1})\right)\subset\pi_{2}\left(\WF(K)_{\supp\phi\otimes\psi}\right) \\[0.2cm]
\Rightarrow\ &\ \pi_{2}\left(\WF(K)_{\supp^{\circ}\phi\otimes\psi}\right)_{|l=0}\subset\Sigma(\phi(\mathcal{K}\psi))\subset\pi_{2}\left(\WF(K)_{\supp\phi\otimes\psi}\right)_{|l=0}.
\end{align}
Letting $\supp\phi\rightarrow\{x_{0}\}$ proves the theorem. \qed
\end{proof}
\end{theorem}
Along similar lines one obtains an extension theorem to the latter
\begin{theorem}[cf. Theorem 8.2.13. \cite{HoermanderTheAnalysisOf1}]
\label{thm:kernelextension}
The exists a unique extension of $\mathcal{K}$ to those \mbox{$u\in\cE'(V)$} with $\WF(u)\cap\left(-\WF(K)_{V|U}\right)=\emptyset$, s.t.
\begin{equation}
\label{eq:kernelextension}
\cE'(M)\cap\cD'_{\Gamma}(V)\ni u\longrightarrow\mathcal{K}u\in\cD'(U)
\end{equation}
is continuous for all compact sets $M\in V$ and closed cones $\Gamma$ with $\Gamma\cap\left(-\WF(K)_{V|U}\right)=\emptyset$. Define
\begin{eqnarray}
\label{eq:primedwavefrontset}
\WF'(K) & = \left\{(x,y;k,l)\ |\ (x,y;k,-l)\in\WF(K)\right\}, \\ \nonumber 
'\WF(K) & = \left\{(x,y;k,l)\ |\ (x,y;-k,l)\in\WF(K)\right\}.
\end{eqnarray}
One has
\begin{equation}
\label{eq:wavefrontsetextendedkernelmap}
\WF(\mathcal{K}u)\subset\WF'(K)\circ\WF(u)\cup\WF(K)_{U|\supp u},
\end{equation}
where $\circ$ denotes the composition, i.e.\footnote{From theorem \ref{thm:distkernelcomposition}, one additionally has $\WF((1\otimes u)K)_{U|\supp^{\circ}u}\subset\WF(\mathcal{K}u)$.}
\begin{equation}
\label{eq:wavefrontsetcomposition}
\WF'(K)\circ\WF(u)=\left\{(x,k)\ |\ (x,y;k,-l)\in\WF(K),\ (y,l)\in\WF(u)\right\}.
\end{equation}
\end{theorem}
and a composition theorem for this type of maps.
\begin{theorem}[cf. Theorem 8.2.14. \cite{HoermanderTheAnalysisOf1}]
\label{thm:distkernelcomposition2}
Let $U\subset\R^{n}$, $V\subset\R^{m}$ and $W\subset\R^{p}$ be open sets and $K_{1}\in\cD'(U\times V)$, $K_{2}\in\cD'(V\times W)$. Furthermore assume the projection
\begin{equation}
\label{eq:propermap}
\pi_{2}:\supp K_{2}\longrightarrow W
\end{equation}
to be proper, i.e. preimages of compact sets are compact, and
\begin{equation}
\label{eq:wavefrontsetintersectionkernel}
\WF'(K_{1})_{V|U}\cap\WF(K_{2})_{V|W}=\emptyset.
\end{equation}
Then the composition $\mathcal{K}_{1}\circ\mathcal{K}_{2}$ is defined and its kernel $K$ satisfies the following condition on its wave front set
\begin{eqnarray}
\label{eq:wavefrontsetkernelcomposition2}
\WF(K) & \subset & \WF'(K_{1})\circ\WF(K_{2}) \\[0.2cm] \nonumber
&  & \cup\left\{(x,z;k,0)\ |\ (x,y;k,0)\in\WF(K_{1}),\ (y,z)\in\supp K_{2}\right\} \\[0.2cm] \nonumber
&  & \cup\left\{(x,z;0,m)\ |\ (x,y)\in\supp K_{1},\ (y,z;0,m)\in\WF(K_{2})\right\} \\[0.2cm]
& \subset & \WF'(K_{1})\circ\WF(K_{2})\cup\left(\WF(K_{1})_{U|\pi_{1}(\supp K_{2})}\times\left(\pi_{2}(\supp K_{2})\times\{0\}\right)\right) \\[0.2cm] \nonumber
&  & \cup\left(\left(\pi_{1}(\supp K_{1})\times\{0\}\right)\times\WF(K_{2})_{W|\pi_{2}(\supp K_{1})}\right) 
\end{eqnarray}
\end{theorem}
Finally, we need a theorem shedding light on the interplay between wave front sets and differential operators (cf. Theorem 8.3.1. \cite{HoermanderTheAnalysisOf1} and Theorem 26.1.1. \cite{HoermanderTheAnalysisOf4}).
\begin{theorem}
\label{thm:characteristicwavefrontset}
Let $P=\sum_{n\leq m}P_{n}(x,\partial_{x})$ be a linear ($C^{\infty}$-)differential operator of order $m$ on a ($C^{\infty}$-)manifold $M$, then
\begin{equation}
\label{eq:characteristicwavefrontset}
\WF(u)\setminus\WF(Pu)\subset\Char P,\ u\in\cD'(M),
\end{equation}
where $\Char P=\left\{(x,k)\in\textup{T*M}\setminus\{0\}\ |\ P_{m}(x,k)=0\right\}$ denotes the \textup{characteristic set} of $P$. \\
If additionally the \textup{principal symbol} $P_{m}$ is real and homogeneous of degree $m$, $\WF(u)\setminus\WF(Pu)$ will be invariant under the flow of the Hamiltonian vector field associated with $P_{m}$ w.r.t. to the natural symplectic structure on $T^{*}M$.
\end{theorem}
\renewcommand\refname{6\ \ References}
\addcontentsline{toc}{section}{6\ \ References}
\label{sec:ref}
\bibliography{microlocalv2.bbl} 

\begin{thebibliography}{10}
\expandafter\ifx\csname url\endcsname\relax
  \def\url#1{\texttt{#1}}\fi
\expandafter\ifx\csname urlprefix\endcsname\relax\def\urlprefix{URL }\fi
\expandafter\ifx\csname href\endcsname\relax
  \def\href#1#2{#2} \def\path#1{#1}\fi

\bibitem{FredenhagenPerturbativeAlgebraicQuantum}
K.~{F}redenhagen, K.~{R}ejzner,
  \href{http://link.springer.com/chapter/10.1007/978-3-319-09949-1_2}{{P}erturbative
  {A}lgebraic {Q}uantum {F}ield {T}heory}, in: D.~Calaque, T.~Strobl (Eds.),
  Mathematical Aspects of Quantum Field Theories, Mathematical Physics Studies,
  Springer, 2015, pp. 17--55.
\newblock \href {http://dx.doi.org/10.1007/978-3-319-09949-1_2}
  {\path{doi:10.1007/978-3-319-09949-1_2}}.
\newline\urlprefix\url{http://link.springer.com/chapter/10.1007/978-3-319-09949-1_2}

\bibitem{BrunettiQuantumGravityFrom}
R.~{B}runetti, K.~{F}redenhagen, K.~{R}ejzner,
  \href{http://arxiv.org/abs/1306.1058}{{Quantum gravity from the point of view
  of locally covariant quantum field theory}}, ar{X}iv:1306.1058.
\newline\urlprefix\url{http://arxiv.org/abs/1306.1058}

\bibitem{WaldQuantumFieldTheory}
R.~M. {W}ald, {Q}uantum {F}ield {T}heory in {C}urved {S}pacetime and {B}lack
  {H}ole {T}hermodynamics, Chicago Lectures in Physics, The University of
  Chicago Press, 1994.

\bibitem{BrunettiMicrolocalAnalysisAnd}
R.~{B}runetti, K.~{F}redenhagen,
  \href{http://link.springer.com/article/10.1007/s002200050004}{{Microlocal
  Analysis and Interacting Quantum Field Theories: Renormalization on Physical
  Backgrounds}}, {C}ommunications in {M}athematical {P}hysics 208~(3) (2000)
  623 -- 661.
\newblock \href {http://dx.doi.org/10.1007/s002200050004}
  {\path{doi:10.1007/s002200050004}}.
\newline\urlprefix\url{http://link.springer.com/article/10.1007/s002200050004}

\bibitem{HollandsExistenceOfLocal}
S.~{H}ollands, R.~M. {W}ald,
  \href{http://link.springer.com/article/10.1007/s00220-002-0719-y}{{E}xistence
  of {L}ocal {C}ovariant {T}ime {O}rdered {P}roducts of {Q}uantum {F}ields in
  {C}urved {S}pacetime}, {C}ommunications in {M}athematical {P}hysics 231~(2)
  (2002) 309--345.
\newblock \href {http://arxiv.org/abs/gr-qc/0111108}
  {\path{arXiv:gr-qc/0111108}}, \href
  {http://dx.doi.org/10.1007/s00220-002-0719-y}
  {\path{doi:10.1007/s00220-002-0719-y}}.
\newline\urlprefix\url{http://link.springer.com/article/10.1007/s00220-002-0719-y}

\bibitem{BrunettiTheGenerallyCovariant}
R.~{B}runetti, K.~{F}redenhagen, R.~{V}erch,
  \href{http://link.springer.com/article/10.1007%2Fs00220-003-0815-7}{{The
  Generally Covariant Locality Principle -- A New Paradigm for Local Quantum
  Physics}}, {C}ommunications in {M}athematical {P}hysics 237~(1-2) (2003)
  31--68.
\newblock \href {http://arxiv.org/abs/math-ph/0112041}
  {\path{arXiv:math-ph/0112041}}, \href
  {http://dx.doi.org/10.1007/s00220-003-0815-7}
  {\path{doi:10.1007/s00220-003-0815-7}}.
\newline\urlprefix\url{http://link.springer.com/article/10.1007%2Fs00220-003-0815-7}

\bibitem{FullingSingularityStructureOf2}
S.~A. {F}ulling, F.~J. {N}arcowich, R.~M. {W}ald,
  \href{http://www.sciencedirect.com/science/article/pii/0003491681900981}{{Singularity
  structure of the two-point function in quantum field theory in curved
  spacetime, II}}, {A}nnals of {P}hysics 136~(2) (1981) 243--272.
\newblock \href {http://dx.doi.org/10.1016/0003-4916(81)90098-1}
  {\path{doi:10.1016/0003-4916(81)90098-1}}.
\newline\urlprefix\url{http://www.sciencedirect.com/science/article/pii/0003491681900981}

\bibitem{JunkerHadamardStatesAdiabatic}
W.~{J}unker,
  \href{http://www.worldscientific.com/doi/abs/10.1142/S0129055X9600041X}{{H}adamard
  states, adiabatic vacua and the construction of physical states for scalar
  quantum fields on curved spacetime}, {R}eviews in {M}athematical {P}hysics
  8~(8) (1996) 1091--1159.
\newblock \href {http://dx.doi.org/10.1142/S0129055X9600041X}
  {\path{doi:10.1142/S0129055X9600041X}}.
\newline\urlprefix\url{http://www.worldscientific.com/doi/abs/10.1142/S0129055X9600041X}

\bibitem{GerardConstructionOfHadamard}
C.~{G}{\'e}rard, M.~{W}rochna,
  \href{http://link.springer.com/article/10.1007/s00220-013-1824-9}{{C}onstruction
  of {H}adamard {S}tates by {P}seudo-{D}ifferential {C}alculus},
  {C}ommunications in {M}athematical {P}hysics 325~(2) (2014) 713--755.
\newblock \href {http://dx.doi.org/10.1007/s00220-013-1824-9}
  {\path{doi:10.1007/s00220-013-1824-9}}.
\newline\urlprefix\url{http://link.springer.com/article/10.1007/s00220-013-1824-9}

\bibitem{RadzikowskiMicrolocalApproachTo}
M.~J. {R}adzikowski,
  \href{http://link.springer.com/article/10.1007/BF02100096}{{M}icro-local
  approach to the {H}adamard condition in quantum field theory on curved
  space-time}, {C}ommunications in {M}athematical {P}hysics 179~(3) (1996)
  529--553.
\newblock \href {http://dx.doi.org/10.1007/BF02100096}
  {\path{doi:10.1007/BF02100096}}.
\newline\urlprefix\url{http://link.springer.com/article/10.1007/BF02100096}

\bibitem{GieselScalarMaterialReference}
K.~{G}iesel, T.~{T}hiemann, \href{http://arxiv.org/abs/1206.3807}{{S}calar
  {M}aterial {R}eference {S}ystems and {L}oop {Q}uantum {G}ravity},
  ar{X}iv:1206.3807.
\newline\urlprefix\url{http://arxiv.org/abs/1206.3807}

\bibitem{LuedersLocalQuasiequivalenceAnd}
C.~{L}{\"u}ders, J.~E. {R}oberts,
  \href{http://link.springer.com/article/10.1007/BF02102088}{{L}ocal
  quasiequivalence and adiabatic vacuum states}, {C}ommunications in
  {M}athematical {P}hysics 134~(1) (1990) 29--63.
\newblock \href {http://dx.doi.org/10.1007/BF02102088}
  {\path{doi:10.1007/BF02102088}}.
\newline\urlprefix\url{http://link.springer.com/article/10.1007/BF02102088}

\bibitem{JunkerAdiabaticVacuumStates}
W.~{J}unker, E.~{S}chrohe,
  \href{http://link.springer.com/article/10.1007/s000230200001}{{A}diabatic
  {V}acuum {S}tates on {G}eneral {S}pacetime {M}anifolds: {D}efinition,
  {C}onstruction, and {P}hysical {P}roperties}, {A}nnales {H}enri
  {P}oincar{\'e} 3~(6) (2002) 1113--1181.
\newblock \href {http://dx.doi.org/10.1007/s000230200001}
  {\path{doi:10.1007/s000230200001}}.
\newline\urlprefix\url{http://link.springer.com/article/10.1007/s000230200001}

\bibitem{AgulloExtensionOfThe}
I.~{A}gullo, A.~{A}shtekar, W.~{N}elson,
  \href{http://prd.aps.org/abstract/PRD/v87/i4/e043507}{{Extension of the
  quantum theory of cosmological perturbations to the Planck era}}, {P}hysical
  {R}eview {D} 87~(4) (2013) 043507.
\newblock \href
  {http://dx.doi.org/http://dx.doi.org/10.1103/PhysRevD.87.043507}
  {\path{doi:http://dx.doi.org/10.1103/PhysRevD.87.043507}}.
\newline\urlprefix\url{http://prd.aps.org/abstract/PRD/v87/i4/e043507}

\bibitem{AgulloThePreInflationary}
I.~{A}gullo, A.~{A}shtekar, W.~{N}elson,
  \href{http://iopscience.iop.org/0264-9381/30/8/085014}{{The pre-inflationary
  dynamics of loop quantum cosmology: confronting quantum gravity with
  observations}}, {C}lassical and {Q}uantum {G}ravity 30~(8) (2013) 085014.
\newblock \href {http://dx.doi.org/10.1088/0264-9381/30/8/085014}
  {\path{doi:10.1088/0264-9381/30/8/085014}}.
\newline\urlprefix\url{http://iopscience.iop.org/0264-9381/30/8/085014}

\bibitem{DimockAlgebrasOfLocal}
J.~{D}imock,
  \href{http://link.springer.com/article/10.1007/BF01269921}{{A}lgebras of
  local observables on a manifold}, {C}ommunications in {M}athematical
  {P}hysics 77~(3) (1980) 219--228.
\newblock \href {http://dx.doi.org/10.1007/BF01269921}
  {\path{doi:10.1007/BF01269921}}.
\newline\urlprefix\url{http://link.springer.com/article/10.1007/BF01269921}

\bibitem{BaerWaveEquationsOn}
C.~{B}{\"a}r, N.~{G}inoux, F.~{P}f{\"a}ffle,
  \href{http://www.ems-ph.org/books/book.php?proj_nr=54&srch=series|esi}{{W}ave
  {E}quations on {L}orentzian {M}anifolds and {Q}uantization}, ESI Lectures in
  Mathematics and Physics, European Mathematical Society Publishing House,
  2007.
\newblock \href {http://dx.doi.org/10.4171/037} {\path{doi:10.4171/037}}.
\newline\urlprefix\url{http://www.ems-ph.org/books/book.php?proj_nr=54&srch=series|esi}

\bibitem{HoermanderFourierIntegralOperators}
L.~{H}{\"o}rmander,
  \href{http://www.springerlink.com/index/t202410l4v37r13m.pdf}{{F}ourier
  integral operators. {I}}, {A}cta mathematica 127~(1) (1971) 79--183.
\newblock \href {http://dx.doi.org/10.1007/BF02392052}
  {\path{doi:10.1007/BF02392052}}.
\newline\urlprefix\url{http://www.springerlink.com/index/t202410l4v37r13m.pdf}

\bibitem{DabrowskiFunctionalPropertiesOf}
Y.~{D}abrowski, C.~{B}rouder,
  \href{http://link.springer.com/article/10.1007/s00220-014-2156-0}{{F}unctional
  {P}roperties of {H}{\"o}rmander's {S}pace of {D}istributions {H}aving a
  {S}pecified {W}avefront {S}et}, {C}ommunications in {M}athematical {P}hysics
  332~(3) (2014) 1345--1380.
\newblock \href {http://dx.doi.org/10.1007/s00220-014-2156-0}
  {\path{doi:10.1007/s00220-014-2156-0}}.
\newline\urlprefix\url{http://link.springer.com/article/10.1007/s00220-014-2156-0}

\bibitem{KrieglTheConvenientSetting}
A.~{K}riegl, P.~W. {M}ichor,
  \href{http://www.mat.univie.ac.at/~michor/apbookh-ams.pdf}{{T}he {C}onvenient
  {S}etting for {G}lobal {A}nalysis}, Vol.~53 of Mathematical Surveys and
  Monographs, American Mathematical Society, 1997.
\newline\urlprefix\url{http://www.mat.univie.ac.at/~michor/apbookh-ams.pdf}

\bibitem{FriedlanderTheWaveEquation}
F.~{F}riedlander, \href{http://www.cambridge.org/9780521205672}{{T}he {W}ave
  {E}quation on a {C}urved {S}pace-{T}ime}, Cambridge Monographs on
  Mathematical Physics, Cambridge University Press, 1975.
\newline\urlprefix\url{http://www.cambridge.org/9780521205672}

\bibitem{BernalSmoothnessOfTime}
A.~N. {B}ernal, M.~{S}{\'a}nchez,
  \href{http://link.springer.com/article/10.1007/s00220-005-1346-1}{{S}moothness
  of {T}ime {F}unctions and the {M}etric {S}plitting of {G}lobally {H}yperbolic
  {S}pacetimes}, {C}ommunications in {M}athematical {P}hysics 257~(1) (2005)
  43--50.
\newblock \href {http://dx.doi.org/10.1007/s00220-005-1346-1}
  {\path{doi:10.1007/s00220-005-1346-1}}.
\newline\urlprefix\url{http://link.springer.com/article/10.1007/s00220-005-1346-1}

\bibitem{BaumNormallyHyperbolicOperators}
H.~{B}aum, I.~{K}ath,
  \href{http://link.springer.com/article/10.1007/BF00129896}{{N}ormally
  hyperbolic operators, the {H}uygens property and conformal geometry},
  {A}nnals of {G}lobal {A}nalysis and {G}eometry 14~(4) (1996) 315--371.
\newblock \href {http://dx.doi.org/10.1007/BF00129896}
  {\path{doi:10.1007/BF00129896}}.
\newline\urlprefix\url{http://link.springer.com/article/10.1007/BF00129896}

\bibitem{HoermanderTheAnalysisOf1}
L.~{H}{\"o}rmander, {T}he {A}nalysis of {L}inear {P}artial {D}ifferential
  {O}perators: {V}ol.: 1.: {D}istribution {T}heory and {F}ourier {A}nalysis,
  Vol. 256 of Grundlehren der mathematischen Wissenschaften, Springer, 1983.

\bibitem{LindbladCounterexamplesToLocal}
H.~{L}indblad, \href{http://www.jstor.org/stable/25098452}{{C}ounterexamples to
  local existence for semi-linear wave equations}, {A}merican {J}ournal of
  {M}athematics 118~(1) (1996) 1--16.
\newblock \href {http://dx.doi.org/10.2307/25098452}
  {\path{doi:10.2307/25098452}}.
\newline\urlprefix\url{http://www.jstor.org/stable/25098452}

\bibitem{DuistermaatFourierIntegralOperators}
J.~J. {D}uistermaat, L.~{H}{\"o}rmander,
  \href{http://www.springerlink.com/index/42171T527J141H45.pdf}{{F}ourier
  integral operators. {II}}, {A}cta mathematica 128~(1) (1972) 183--269.
\newblock \href {http://dx.doi.org/10.1007/BF02392165}
  {\path{doi:10.1007/BF02392165}}.
\newline\urlprefix\url{http://www.springerlink.com/index/42171T527J141H45.pdf}

\bibitem{BuchholzTheResolventAlgebra1}
D.~{B}uchholz, H.~{G}rundling,
  \href{http://www.sciencedirect.com/science/article/pii/S0022123608000694}{{T}he
  resolvent algebra: {A} new approach to canonical quantum systems}, {J}ournal
  of {F}unctional {A}nalysis 254~(11) (2008) 2725--2779.
\newblock \href {http://dx.doi.org/10.1016/j.jfa.2008.02.011}
  {\path{doi:10.1016/j.jfa.2008.02.011}}.
\newline\urlprefix\url{http://www.sciencedirect.com/science/article/pii/S0022123608000694}

\bibitem{BuchholzTheResolventAlgebra2}
D.~{B}uchholz,
  \href{http://www.sciencedirect.com/science/article/pii/S0022123613004382}{{T}he
  resolvent algebra: {I}deals and dimension}, {J}ournal of {F}unctional
  {A}nalysis 266~(5) (2014) 3286--3302.
\newline\urlprefix\url{http://www.sciencedirect.com/science/article/pii/S0022123613004382}

\bibitem{BratteliOperatorAlgebrasAnd1}
O.~{B}ratteli, D.~W. {R}obinson, {O}perator {A}lgebras and {Q}uantum
  {S}tatistical {M}echanics 1: ${C}^{*}$-and ${W}^{*}$-{A}lgebras, {S}ymmetry
  {G}roups, {D}ecomposition of {S}tates, 2nd Edition, Texts and Monographs in
  Physics, Springer, 1987.

\bibitem{BratteliOperatorAlgebrasAnd2}
O.~{B}ratteli, D.~W. {R}obinson, {O}perator {A}lgebras and {Q}uantum
  {S}tatistical {M}echanics 2: {E}quilibrium {S}tates, {M}odels in {Q}uantum
  {S}tatistical {M}echanics, 2nd Edition, Texts and Monographs in Physics,
  Springer, 1997.

\bibitem{BrunettiTheMicrolocalSpectrum}
R.~{B}runetti, K.~{F}redenhagen, M.~{K}{\"o}hler,
  \href{http://link.springer.com/article/10.1007/BF02099626}{{The microlocal
  spectrum condition and Wick polynomials of free fields on curved
  spacetimes}}, {C}ommunications in {M}athematical {P}hysics 180~(3) (1996)
  633--652.
\newblock \href {http://arxiv.org/abs/gr-qc/9510056}
  {\path{arXiv:gr-qc/9510056}}, \href {http://dx.doi.org/10.1007/BF02099626}
  {\path{doi:10.1007/BF02099626}}.
\newline\urlprefix\url{http://link.springer.com/article/10.1007/BF02099626}

\bibitem{KratzertSingularityStructureOf}
K.~{K}ratzert,
  \href{ftp://it-ftp1.desy.de/pub/preprints/desy/thesis/desy-thesis-99-020.ps.gz}{{S}ingularit{\"a}tsstruktur
  der {Z}weipunkfunktion des freien {D}iracfeldes in einer global
  hyperbolischen {R}aumzeit}, Master's thesis, DESY (January 1999).
\newline\urlprefix\url{ftp://it-ftp1.desy.de/pub/preprints/desy/thesis/desy-thesis-99-020.ps.gz}

\bibitem{HollandsTheHadamardCondition}
S.~{H}ollands,
  \href{http://link.springer.com/article/10.1007/s002200000350}{{T}he
  {H}adamard {C}ondition for {D}irac {F}ields and {A}diabatic {S}tates on
  {R}obertson--{W}alker {S}pacetimes}, {C}ommunications in {M}athematical
  {P}hysics 216~(3) (2001) 635--661.
\newblock \href {http://dx.doi.org/10.1007/s002200000350}
  {\path{doi:10.1007/s002200000350}}.
\newline\urlprefix\url{http://link.springer.com/article/10.1007/s002200000350}

\bibitem{DAntoniNuclearityLocalQuasiequivalence}
C.~{D}'{A}ntoni, S.~{H}ollands,
  \href{http://link.springer.com/article/10.1007/s00220-005-1398-2}{{N}uclearity,
  {L}ocal {Q}uasiequivalence and {S}plit {P}roperty for {D}irac {Q}uantum
  {F}ields in {C}urved {S}pacetime}, {C}ommunications in {M}athematical
  {P}hysics 261~(1) (2006) 133--159.
\newblock \href {http://dx.doi.org/10.1007/s00220-005-1398-2}
  {\path{doi:10.1007/s00220-005-1398-2}}.
\newline\urlprefix\url{http://link.springer.com/article/10.1007/s00220-005-1398-2}

\bibitem{DappiaggiTheExtendedAlgebra}
C.~{D}appiaggi, T.-P. {H}ack, N.~{P}inamonti,
  \href{http://www.worldscientific.com/doi/abs/10.1142/S0129055X09003864}{{T}he
  extended algebra of observables for {D}irac fields and the trace anomaly of
  their stress-energy tensor}, {R}eviews in {M}athematical {P}hysics 21~(10)
  (2009) 1241--1312.
\newblock \href {http://dx.doi.org/10.1142/S0129055X09003864}
  {\path{doi:10.1142/S0129055X09003864}}.
\newline\urlprefix\url{http://www.worldscientific.com/doi/abs/10.1142/S0129055X09003864}

\bibitem{SandersTheLocallyCovariant}
K.~{S}anders,
  \href{http://www.worldscientific.com/doi/abs/10.1142/S0129055X10003990}{{T}he
  locally covariant {D}irac field}, {R}eviews in {M}athematical {P}hysics
  22~(04) (2010) 381--430.
\newblock \href {http://dx.doi.org/10.1142/S0129055X10003990}
  {\path{doi:10.1142/S0129055X10003990}}.
\newline\urlprefix\url{http://www.worldscientific.com/doi/abs/10.1142/S0129055X10003990}

\bibitem{RejznerFermionicFieldsIn}
K.~{R}ejzner,
  \href{http://www.worldscientific.com/doi/abs/10.1142/S0129055X11004503}{{F}ermionic
  fields in the functional approach to classical field theory}, {R}eviews in
  {M}athematical {P}hysics 23~(09) (2011) 1009--1033.
\newblock \href {http://dx.doi.org/10.1142/S0129055X11004503}
  {\path{doi:10.1142/S0129055X11004503}}.
\newline\urlprefix\url{http://www.worldscientific.com/doi/abs/10.1142/S0129055X11004503}

\bibitem{HollandsRenormalizedQuantumYangMills}
S.~{H}ollands,
  \href{http://www.worldscientific.com/doi/abs/10.1142/S0129055X08003420}{{Renormalized
  Quantum Yang--Mills Fields in Curved Spacetime}}, {R}eviews in {M}athematical
  {P}hysics 20~(09) (2008) 1033--1172.
\newblock \href {http://dx.doi.org/10.1142/S0129055X08003420}
  {\path{doi:10.1142/S0129055X08003420}}.
\newline\urlprefix\url{http://www.worldscientific.com/doi/abs/10.1142/S0129055X08003420}

\bibitem{FredenhagenBatalinVilkoviskyFormalism1}
K.~{F}redenhagen, K.~{R}ejzner,
  \href{http://link.springer.com/article/10.1007/s00220-012-1487-y}{{B}atalin-{V}ilkovisky
  {F}ormalism in the {F}unctional {A}pproach to {C}lassical {F}ield {T}heory},
  {C}ommunications in {M}athematical {P}hysics 314~(1) (2012) 93--127.
\newblock \href {http://dx.doi.org/10.1007/s00220-012-1487-y}
  {\path{doi:10.1007/s00220-012-1487-y}}.
\newline\urlprefix\url{http://link.springer.com/article/10.1007/s00220-012-1487-y}

\bibitem{FredenhagenBatalinVilkoviskyFormalism2}
K.~{F}redenhagen, K.~{R}ejzner,
  \href{http://link.springer.com/article/10.1007/s00220-012-1601-1}{{B}atalin-{V}ilkovisky
  {F}ormalism in {P}erturbative {A}lgebraic {Q}uantum {F}ield {T}heory},
  {C}ommunications in {M}athematical {P}hysics 317~(3) (2013) 697--725.
\newblock \href {http://dx.doi.org/10.1007/s00220-012-1601-1}
  {\path{doi:10.1007/s00220-012-1601-1}}.
\newline\urlprefix\url{http://link.springer.com/article/10.1007/s00220-012-1601-1}

\bibitem{FredenhagenLocalCovarianceAnd}
K.~{F}redenhagen, K.~{R}ejzner,
  \href{http://link.springer.com/chapter/10.1007/978-3-0348-0043-3_2}{{L}ocal
  {C}ovariance and {B}ackground {I}ndependence}, in: F.~Finster, O.~M{\"u}ller,
  M.~Nardmann, J.~Tolksdorf, E.~Zeidler (Eds.), Quantum Field Theory and
  Gravity, Springer, 2012, Ch.~2, pp. 15--23.
\newblock \href {http://dx.doi.org/10.1007/978-3-0348-0043-3_2}
  {\path{doi:10.1007/978-3-0348-0043-3_2}}.
\newline\urlprefix\url{http://link.springer.com/chapter/10.1007/978-3-0348-0043-3_2}

\bibitem{ONeillSemiRiemannianGeometry}
{B}arrett {O}'{N}eill, {S}emi-{R}iemannian {G}eometry - {W}ith {A}pplications
  to {R}elativity, Pure and Applied Mathematics, Academic Press, 1983.

\bibitem{FullingAspectsOfQuantum}
S.~A. {F}ulling, {Aspects of Quantum Field Theory in Curved Space-Time}, no.~17
  in London Mathematical Society Student Texts, Cambridge University Press,
  1989.

\bibitem{ChernoffEssentialSelfAdjointness}
P.~R. {C}hernoff,
  \href{http://dx.doi.org/10.1016/0022-1236(73)90003-7}{{E}ssential
  self-adjointness of powers of generators of hyperbolic equations}, {J}ournal
  of {F}unctional {A}nalysis 12~(4) (1973) 401--414.
\newblock \href {http://dx.doi.org/10.1016/0022-1236(73)90003-7}
  {\path{doi:10.1016/0022-1236(73)90003-7}}.
\newline\urlprefix\url{http://dx.doi.org/10.1016/0022-1236(73)90003-7}

\bibitem{HoermanderTheAnalysisOf3}
L.~{H}{\"o}rmander, {T}he {A}nalysis of {L}inear {P}artial {D}ifferential
  {O}perators: {V}ol.: 3.: {P}seudo-{D}ifferential {O}perators, Vol. 274 of
  Grundlehren der mathematischen Wissenschaften, Springer, 1985.

\bibitem{GelfandGeneralizedFunctions4}
I.~M. {G}el'fand, N.~Y. {V}ilenkin, {Generalized Functions, Vol.: 4:
  Applications of Harmonic Analysis}, Academic Press, 1964.

\bibitem{DunfordLinearOperators2}
N.~{D}unford, J.~T. {S}chwartz, {L}inear {O}perators, {P}art {II}: {S}pectral
  {T}heory, {S}elf-{A}djoint {O}perators in {H}ilbert {S}pace, Vol.~7 of Pure
  and Applied Mathematics, Wiley, 1963.

\bibitem{AfshordiADistinguishedVacuum}
N.~{A}fshordi, S.~{A}slanbeigi, R.~D. {S}orkin,
  \href{http://link.springer.com/article/10.1007/JHEP08%282012%29137}{{A}
  distinguished vacuum state for a quantum field in a curved spacetime:
  formalism, features, and {C}osmology}, {J}ournal of {H}igh {E}nergy {P}hysics
  1208~(137).
\newblock \href {http://arxiv.org/abs/1205.1296} {\path{arXiv:1205.1296}},
  \href {http://dx.doi.org/10.1007/JHEP08(2012)137}
  {\path{doi:10.1007/JHEP08(2012)137}}.
\newline\urlprefix\url{http://link.springer.com/article/10.1007/JHEP08%282012%29137}

\bibitem{FewsterOnARecent}
C.~J. {F}ewster, R.~{V}erch,
  \href{http://iopscience.iop.org/0264-9381/29/20/205017/}{{O}n a recent
  construction of 'vacuum-like' quantum field states in curved spacetime},
  {C}lassical and {Q}uantum {G}ravity 29~(20) (2012) 205017.
\newblock \href {http://arxiv.org/abs/1206.1562} {\path{arXiv:1206.1562}},
  \href {http://dx.doi.org/10.1088/0264-9381/29/20/205017}
  {\path{doi:10.1088/0264-9381/29/20/205017}}.
\newline\urlprefix\url{http://iopscience.iop.org/0264-9381/29/20/205017/}

\bibitem{BrumVacuumLikeHadamard}
M.~{B}rum, K.~{F}redenhagen,
  \href{http://iopscience.iop.org/0264-9381/31/2/025024}{'{V}acuum-like'
  {H}adamard states for quantum fields on curved spacetimes}, {C}lassical and
  {Q}uantum {G}ravity 31~(2) (2014) 025024.
\newblock \href {http://arxiv.org/abs/1307.0482} {\path{arXiv:1307.0482}},
  \href {http://dx.doi.org/10.1088/0264-9381/31/2/025024}
  {\path{doi:10.1088/0264-9381/31/2/025024}}.
\newline\urlprefix\url{http://iopscience.iop.org/0264-9381/31/2/025024}

\bibitem{ThiemannKinematicalHilbertSpaces}
T.~{T}hiemann, \href{http://iopscience.iop.org/0264-9381/15/6/006}{{Kinematical
  Hilbert spaces for fermionic and Higgs quantum field theories}}, {C}lassical
  and {Q}uantum {G}ravity 15~(6) (1998) 1487.
\newblock \href {http://dx.doi.org/10.1088/0264-9381/15/6/006}
  {\path{doi:10.1088/0264-9381/15/6/006}}.
\newline\urlprefix\url{http://iopscience.iop.org/0264-9381/15/6/006}

\bibitem{ThiemannQuantumSpinDynamics5}
T.~{T}hiemann, \href{http://iopscience.iop.org/0264-9381/15/5/012}{{Q}uantum
  spin dynamics ({QSD}): {V}. {Q}uantum gravity as the natural regulator of the
  {H}amiltonian constraint of matter quantum field theories}, {C}lassical and
  {Q}uantum {G}ravity 15~(5) (1998) 1281.
\newblock \href {http://dx.doi.org/10.1088/0264-9381/15/5/012}
  {\path{doi:10.1088/0264-9381/15/5/012}}.
\newline\urlprefix\url{http://iopscience.iop.org/0264-9381/15/5/012}

\bibitem{AshtekarPolymerAndFock}
A.~{A}shtekar, J.~{L}ewandowski, H.~{S}ahlmann,
  \href{http://iopscience.iop.org/0264-9381/20/1/103}{{P}olymer and {F}ock
  representations for a scalar field}, {C}lassical and {Q}uantum {G}ravity
  20~(1) (2003) L11.
\newblock \href {http://dx.doi.org/10.1088/0264-9381/20/1/103}
  {\path{doi:10.1088/0264-9381/20/1/103}}.
\newline\urlprefix\url{http://iopscience.iop.org/0264-9381/20/1/103}

\bibitem{ThiemannModernCanonicalQuantum}
T.~{T}hiemann, \href{http://www.cambridge.org/9780521741873}{{M}odern
  {C}anonical {Q}uantum {G}eneral {R}elativity}, Cambridge Monographs on
  Mathematical Physics, Cambridge University Press, 2008.
\newline\urlprefix\url{http://www.cambridge.org/9780521741873}

\bibitem{AshtekarLoopQuantumCosmology}
A.~{A}shtekar, P.~{S}ingh,
  \href{http://iopscience.iop.org/0264-9381/28/21/213001/}{{L}oop quantum
  cosmology: a status report}, {C}lassical and {Q}uantum {G}ravity 28~(21)
  (2011) 213001.
\newblock \href {http://arxiv.org/abs/1108.0893} {\path{arXiv:1108.0893}},
  \href {http://dx.doi.org/10.1088/0264-9381/28/21/213001}
  {\path{doi:10.1088/0264-9381/28/21/213001}}.
\newline\urlprefix\url{http://iopscience.iop.org/0264-9381/28/21/213001/}

\bibitem{VaradarajanFockRepresentationsFrom}
M.~{V}aradarajan,
  \href{http://journals.aps.org/prd/abstract/10.1103/PhysRevD.61.104001}{{F}ock
  representations from \textbf{$U(1)$} holonomy algebras}, {P}hysical {R}eview
  {D} 61~(10) (2000) 104001.
\newblock \href {http://dx.doi.org/10.1103/PhysRevD.61.104001}
  {\path{doi:10.1103/PhysRevD.61.104001}}.
\newline\urlprefix\url{http://journals.aps.org/prd/abstract/10.1103/PhysRevD.61.104001}

\bibitem{VaradarajanPhotonsFromQuantized}
M.~{V}aradarajan,
  \href{http://journals.aps.org/prd/abstract/10.1103/PhysRevD.64.104003}{{P}hotons
  from quantized electric flux representations}, {P}hysical {R}eview {D}
  64~(10) (2001) 104003.
\newblock \href {http://dx.doi.org/10.1103/PhysRevD.64.104003}
  {\path{doi:10.1103/PhysRevD.64.104003}}.
\newline\urlprefix\url{http://journals.aps.org/prd/abstract/10.1103/PhysRevD.64.104003}

\bibitem{SahlmannTowardsTheQFT1}
H.~{S}ahlmann, T.~{T}hiemann,
  \href{http://iopscience.iop.org/0264-9381/23/3/019}{{T}owards the {QFT} on
  curved spacetime limit of {QGR}: {I}. {A} general scheme}, {C}lassical and
  {Q}uantum {G}ravity 23~(3) (2006) 867--908.
\newblock \href {http://dx.doi.org/10.1088/0264-9381/23/3/019}
  {\path{doi:10.1088/0264-9381/23/3/019}}.
\newline\urlprefix\url{http://iopscience.iop.org/0264-9381/23/3/019}

\bibitem{SahlmannTowardsTheQFT2}
H.~{S}ahlmann, T.~{T}hiemann,
  \href{http://iopscience.iop.org/0264-9381/23/3/020}{{T}owards the {QFT} on
  curved spacetime limit of {QGR}: {II}. {A} concrete implementation},
  {C}lassical and {Q}uantum {G}ravity 23~(3) (2006) 909--954.
\newblock \href {http://dx.doi.org/10.1088/0264-9381/23/3/020}
  {\path{doi:10.1088/0264-9381/23/3/020}}.
\newline\urlprefix\url{http://iopscience.iop.org/0264-9381/23/3/020}

\bibitem{GieselBornOppenheimerDecomposition}
K.~{G}iesel, J.~{T}ambornino, T.~{T}hiemann,
  \href{http://arxiv.org/abs/0911.5331}{{Born--Oppenheimer decomposition for
  quantum fields on quantum spacetimes}}, ar{X}iv:0911.5331.
\newline\urlprefix\url{http://arxiv.org/abs/0911.5331}

\bibitem{StottmeisterCoherentStatesQuantumI}
A.~{S}tottmeister, T.~{T}hiemann, {Coherent states, quantum gravity and the
  Born-Oppenheimer approximation, I: General considerations}.

\bibitem{StottmeisterCoherentStatesQuantumII}
A.~{S}tottmeister, T.~{T}hiemann, {Coherent states, quantum gravity and the
  Born-Oppenheimer approximation, II: Compact Lie Groups}.

\bibitem{StottmeisterCoherentStatesQuantumIII}
A.~{S}tottmeister, T.~{T}hiemann, {Coherent states, quantum gravity and the
  Born-Oppenheimer approximation, III: Applications to loop quantum gravity}.

\bibitem{HollandsLocalWickPolynomials}
S.~{H}ollands, R.~M. {W}ald,
  \href{http://link.springer.com/article/10.1007%2Fs002200100540}{{L}ocal
  {W}ick {P}olynomials and {T}ime {O}rdered {P}roducts of {Q}uantum {F}ields in
  {C}urved {S}pacetime}, {C}ommunications in {M}athematical {P}hysics 223~(2)
  (2001) 289--326.
\newblock \href {http://arxiv.org/abs/gr-qc/0103074}
  {\path{arXiv:gr-qc/0103074}}, \href {http://dx.doi.org/10.1007/s002200100540}
  {\path{doi:10.1007/s002200100540}}.
\newline\urlprefix\url{http://link.springer.com/article/10.1007%2Fs002200100540}

\bibitem{SahlmannMicrolocalSpectrumCondition}
H.~{S}ahlmann, R.~{V}erch,
  \href{http://www.worldscientific.com/doi/abs/10.1142/S0129055X01001010}{{M}icrolocal
  spectrum condition and {H}adamard form for vector-valued quantum fields in
  curved spacetime}, {R}eviews in {M}athematical {P}hysics 13~(10) (2001)
  1203--1246.
\newblock \href {http://arxiv.org/abs/math-ph/0008029}
  {\path{arXiv:math-ph/0008029}}, \href
  {http://dx.doi.org/10.1142/S0129055X01001010}
  {\path{doi:10.1142/S0129055X01001010}}.
\newline\urlprefix\url{http://www.worldscientific.com/doi/abs/10.1142/S0129055X01001010}

\bibitem{RudinFunctionalAnalysis}
W.~{R}udin, {F}unctional {A}nalysis, Higher mathematics series, McGraw-Hill,
  1973.

\bibitem{WernerFunktionalanalysis}
D.~{W}erner, {F}unktionalanalysis, 6th Edition, Springer, 1995.

\bibitem{RobertsonTopologicalVectorSpaces}
A.~P. {R}obertson, W.~{R}obertson, {T}opological {V}ector {S}paces, no.~53 in
  Cambridge Texts in Mathematics and Mathematical Physics, Cambridge University
  Press, 1964.

\bibitem{HoermanderTheAnalysisOf4}
L.~{H}{\"o}rmander, {T}he {A}nalysis of {L}inear {P}artial {D}ifferential
  {O}perators: {V}ol.: 4.: {F}ourier {I}ntegral {O}perators, Vol. 275 of
  Grundlehren der mathematischen Wissenschaften, Springer, 1985.

\end{thebibliography}
\bibliographystyle{elsarticle-num}

\end{document}